\documentclass[showpacs,prb,twocolumn,floats,superscriptaddress]{revtex4}

\usepackage{graphicx}
\usepackage{amsmath}
\usepackage{amssymb}
\usepackage{lmodern}
\usepackage[T1]{fontenc}
\usepackage[colorlinks=true,citecolor=blue,linkcolor=blue]{hyperref}
\usepackage{amsfonts}
\usepackage[latin9]{inputenc}
\usepackage{color}
\usepackage{textcomp}
\usepackage{multirow}
\usepackage{makecell}

\renewcommand{\vec}[1]{\ensuremath{\boldsymbol{#1}}}

\begin{document}

\title{Electron collimation at van der Waals domain walls in bilayer graphene}
\date{\today }
\author{H. M. Abdullah}
\email{alshehab211@gmail.com}
\affiliation{Department of Physics, King Fahd University of Petroleum and Minerals, 31261 Dhahran, Saudi Arabia}
\affiliation{Saudi Center for Theoretical Physics, P.O. Box 32741, Jeddah 21438, Saudi Arabia}
\affiliation{Department of Physics, University of Antwerp, Groenenborgerlaan 171, B-2020 Antwerp, Belgium}

\author{D. R. da Costa}
\email{diego$_$rabelo@fisica.ufc.br}
\affiliation{Departamento de Física, Universidade Federal do Ceará, Caixa Postal 6030, Campus do Pici, 60455-900 Fortaleza, Ceará, Brazil}

\author{H. Bahlouli}
\affiliation{Department of Physics, King Fahd University of Petroleum and Minerals, 31261 Dhahran, Saudi Arabia}
\affiliation{Saudi Center for Theoretical Physics, P.O. Box 32741, Jeddah 21438, Saudi Arabia}

\author{A. Chaves}
\affiliation{Departamento de Física, Universidade Federal do Ceará, Caixa Postal 6030, Campus do Pici, 60455-900 Fortaleza, Ceará, Brazil}

\author{F. M. Peeters}
\affiliation{Department of Physics, University of Antwerp, Groenenborgerlaan 171, B-2020 Antwerp, Belgium}

\author{B. Van Duppen}
\email{ben.vanduppen@uantwerpen.be}
\affiliation{Department of Physics, University of Antwerp, Groenenborgerlaan 171, B-2020 Antwerp, Belgium}


\begin{abstract}
We show that a domain wall separating single layer graphene (SLG) and AA-stacked bilayer graphene (AA-BLG) can be used to generate highly collimated electron beams which can be steered by a magnetic field.  Two distinct configurations are studied, namely, locally delaminated AA-BLG and terminated AA-BLG whose terminal edge-type are assumed to be either zigzag or armchair. We investigate the electron scattering  using  semi-classical dynamics and verify the results independently with  wave-packed dynamics simulations. We find that the proposed system supports two distinct types of collimated beams   that correspond to the lower and upper cones in AA-BLG.  Our computational results also reveal that collimation is robust against the number of layers connected to  AA-BLG and terminal edges. \end{abstract}

\maketitle

\section{Introduction}

In the absence of scattering, the wave nature of electrons results in the  analogy between optical and electronic transport\cite{Cheianov2007,Banszerus2016,Wang2018a}. This analogy has provided many novel phenomena in solid-state two-dimensional
electron systems such as lenses\cite{Sivan1990}, beam
splitters\cite{Oliver1999}, and wave guides\cite{Hartmann2010,Williams2011}. In conventional \textit{np} junctions,  optic-like manipulation of electron beams is hindered by the poor electron transmitters. However, in graphene \cite{Novoselov_2004,Geim_2007} electron transmission is enhanced  due to Klein tunneling\cite{Beenakker2008,Klein_1929,Stander2009,Katsnelson2006,Gutierrez2016,Abdullah2018a}. Moreover, its energy spectrum resembles that of photons  which allows experimentalists to use graphene as a test bed for optic-like electron behaviors. For example, two experiments were conducted recently  where a negative refraction was observed for Dirac fermions in graphene\cite{Lee2015} and  the angle-dependent transmission coefficient was  simultaneously measured\cite{Chen2016}. The negative refraction index in graphene  was predicted earlier\cite{Cheianov2007}  where it was found that electrons passing through  \textit{np} junction at specific energy converge on the other side  at the focal point. This behavior is the analog of  a Veselago lens\cite{Veselago1968} that was realized earlier in photonic crystals\cite{Parimi2004,Cubukcu2003} and metamaterials\cite{Song2018,Houck2003,Grbic2004}. These findings led to profound theoretical investigations of electron focusing in SLG\cite{Aidala2007,LaGasse2017,Zhang2018} as well as in AA-\cite{Sanderson_2013} and AB-BLG\cite{Peterfalvi2012}  where a valley selective electronic Veselago lens was proposed.
  
Another analogue to light rays across an optical boundary is the collimation of electrons  across \textit{np} junction.  This analog becomes perfect in the absence of scattering; however, the disorder-induced scattering has precluded the implementation of such an idea. Different proposals have been introduced to maintain collimation of an electron beam such as  using graphene superlattices with  periodic\cite{Park2008} or  disordered\cite{Choi2014} potentials. Another route was also  established by introducing   a  mechanical deformation  to form a    parabolic \textit{pn} junction\cite{Liu2017a} or  carving pinhole
slits in hexagonal Boron Nitride (hBN) encapsulated graphene\cite{Barnard2017}  as well as creating zigzag side contacts \cite{Bhandari2018}. 

Motivated by the recent experiments where a point source of current in single layer graphene \cite{Handschin2015,Kinikar2017} and bilayer\cite{Overweg2017} were achieved, we propose a new system to obtain a highly collimated electron beam which can  be used, for example,  in     Dirac fermion microscope\cite{Boggild2017}. We consider a junction   composed of single layer graphene and AA-BLG. Such system can exist in two configurations where delaminated bilayer graphene or single layer graphene  are connected into AA-BLG  as shown in Figs. \ref{fig:Lattic_Spectrum}(e) and (f), respectively. Recently, it was shown that  such systems exhibit  distinct electronic properties\cite{Abdullah2017,Abdullah2018,Abdullah2018b,Abdullah_2016,Lane2018,Mirzakhani2016}. \textcolor{blue}{} In the   low-energy regime, the Fermi circle in delaminated region is much smaller than its counterpart in the AA-BLG.   This results in a  small refraction index forcing  the transmitted electrons to nearly move in the same  direction. 

In this Article we calculate and compare the collimation of divergent electron beams using two distinct formalisms. In the first approach, we combine in a semi-classical (SC)\cite{Reijnders2013,Milovanovic2015,Peterfalvi2012,Reijnders2017,Milovanovic2014} way quantum mechanical calculation of the transmission  probabilities at a domain wall with a wave propagation described as an optical analog. In the second approach we calculate the  wave-packet dynamics (WD)\cite{Choi2014,Park2008,Maksimova2008,Chaves2010,Krueckl2009} of electrons incident on a domain wall to obtain the carriers trajectories.   To control the direction of the collimated beam, we used a magnetic field to steer the electron beam.  In the first configuration, we assume that   a point source is located in the delaminated bilayer graphene and electrons are emitted  and transmitting into AA-BLG. We find that electrons belonging to the lower and upper cones, within a specific energy range, are bent in    diametrically opposite directions.  This is a manifestation of the fact that the lower cone corresponds to electron-like particles while the upper cone acts as a dispersion of hole-like particles. 

We also show the collimation in the second configuration where single layer graphene is  connected to AA-BLG with zigzag or armchair edges as depicted in Fig. \ref{fig:Lattic_Spectrum}(f). Armchair and zigzag are the two types of edges which are most frequently considered in the study of graphene and BLG samples, although other types of terminations exist due to edge reconstruction. We found that the collimation is robust against the edge shape and the number of layers connected to AA-BLG and we found that the same collimation effects persist.

This paper is organized as follows. In Sec. \ref{Model} we describe the proposed system and present the model. Sec. \ref{Results} is devoted to  numerical results and discussions of collimation and comparison of the two approaches SC and WD. Finally, we conclude by stressing our main findings in Sec. \ref{Concl}.

\section{Model} \label{Model}
\subsection{Atomic structures and boundary conditions}
\begin{figure}[t!]
\vspace{0.cm} 
\centering\graphicspath{{./Figures/}}
\includegraphics[width=1.5  in]{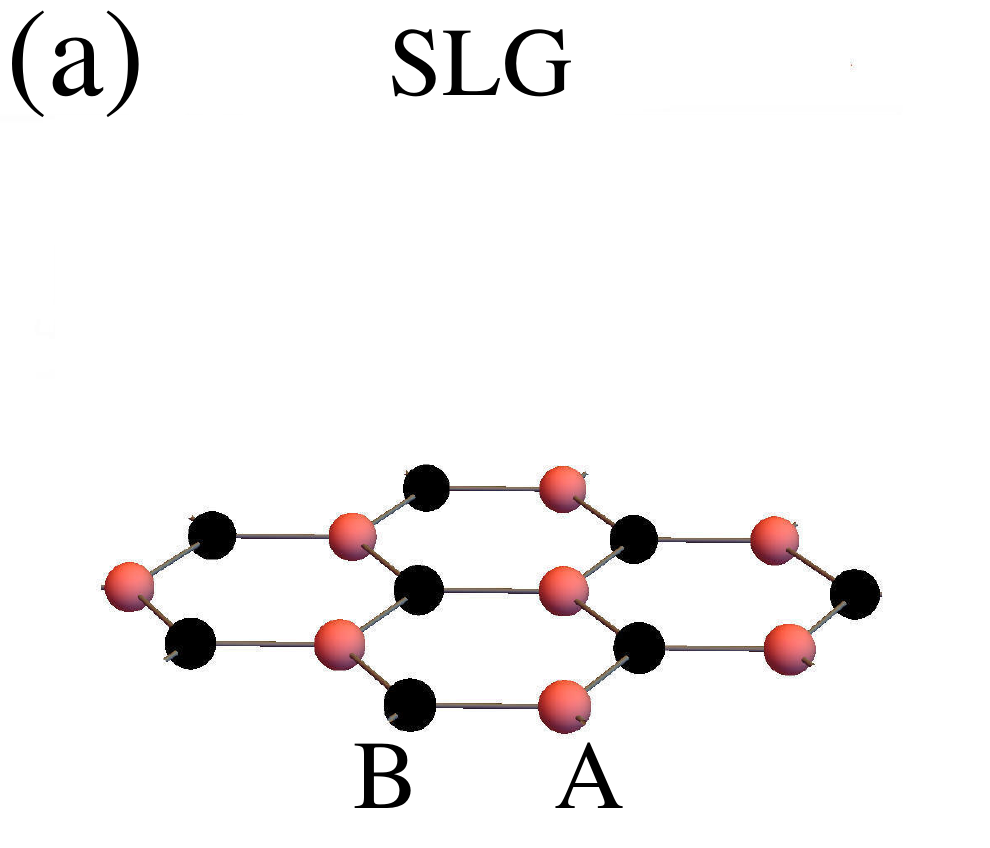}\
\includegraphics[width=1.5  in]{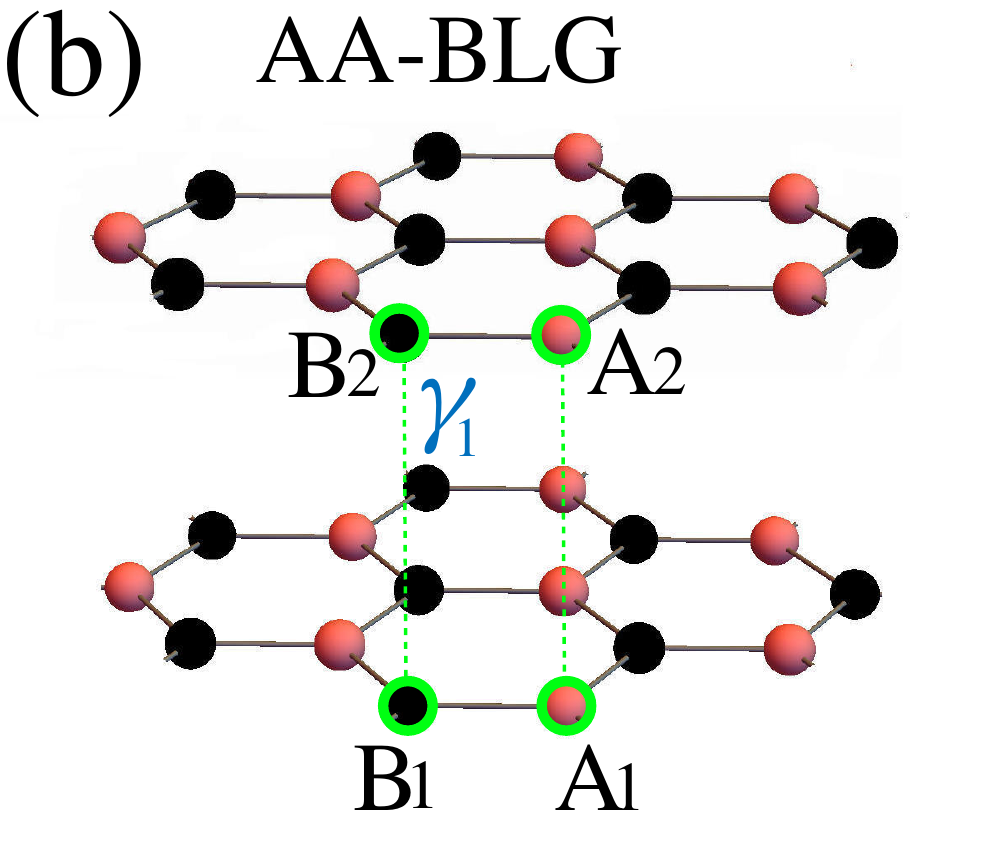}\\
\includegraphics[width=1.5  in]{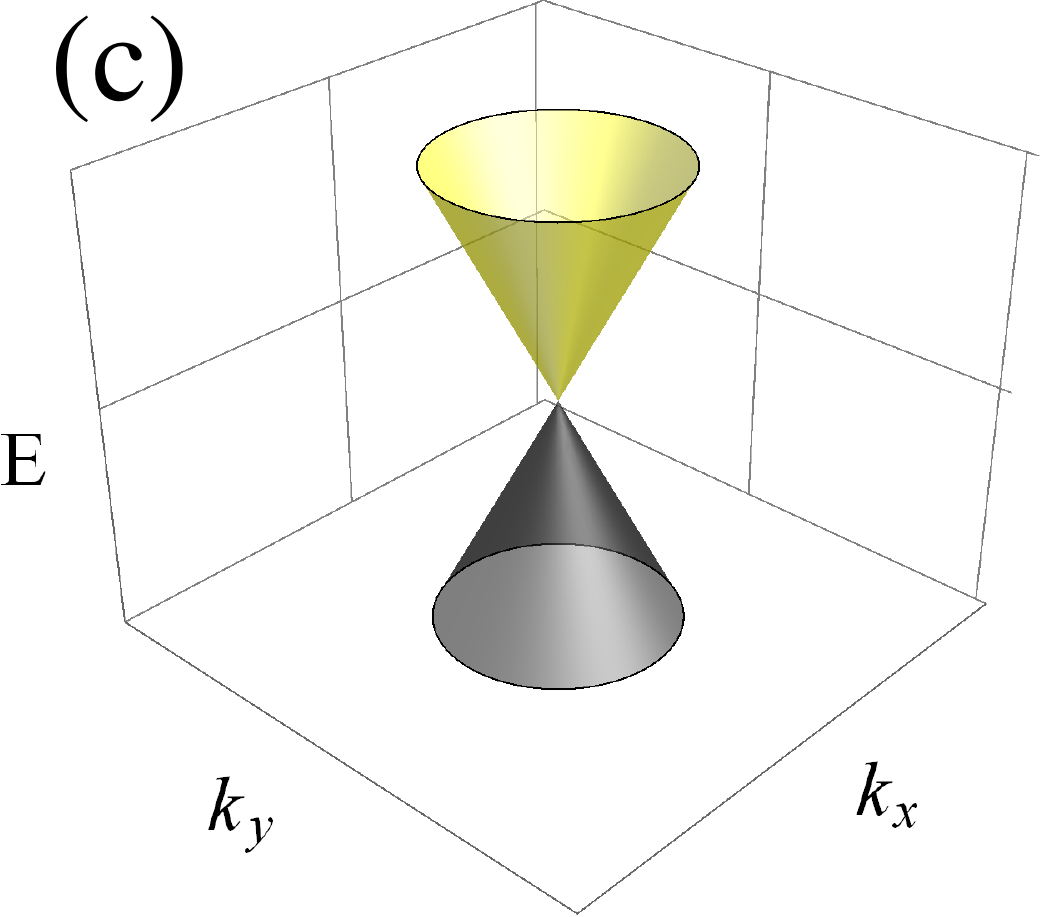}\
\includegraphics[width=1.5  in]{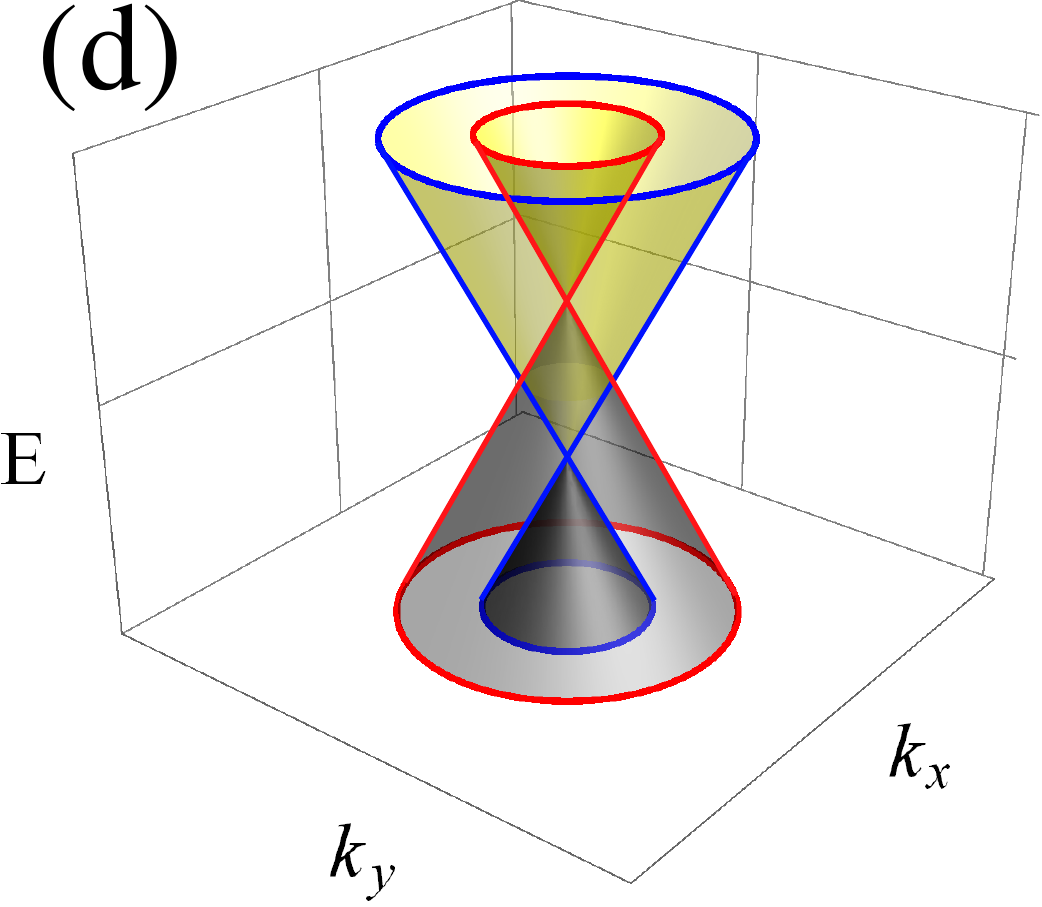}\
\includegraphics[width=3  in]{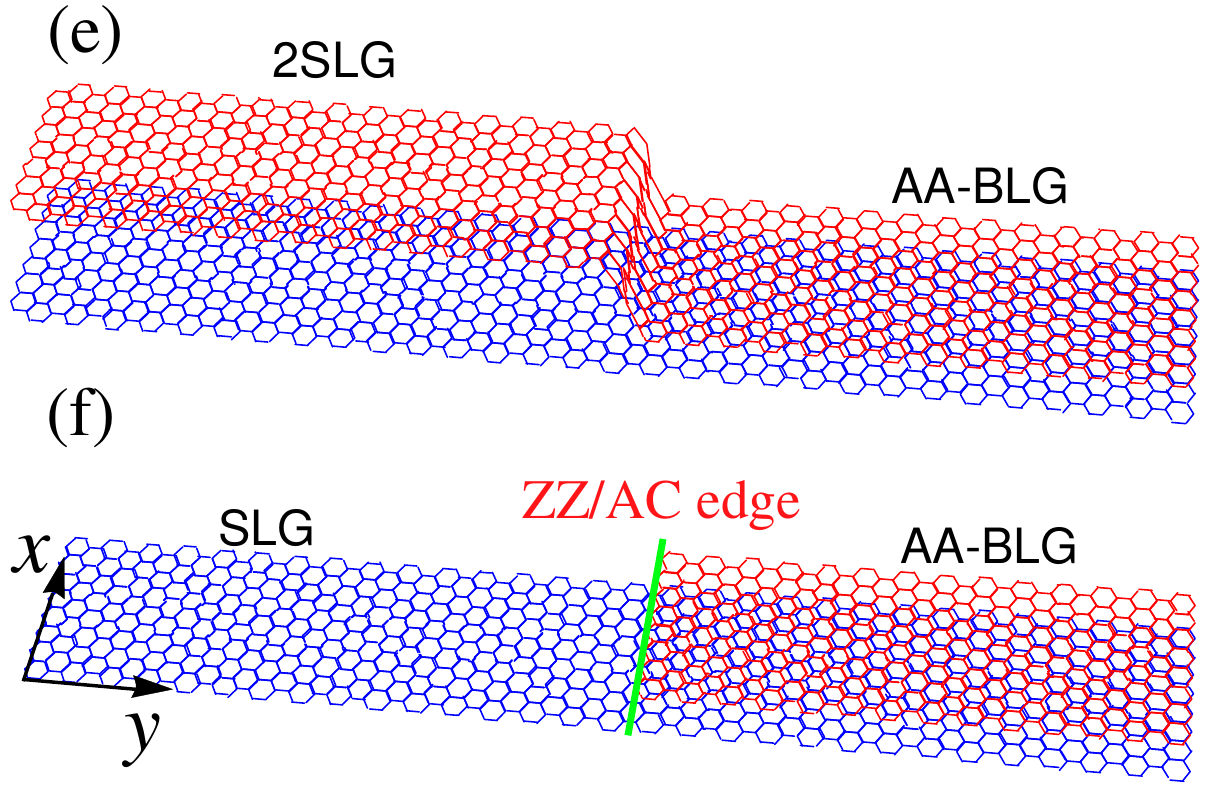}
\vspace{0.cm}
\caption{(Color online) Lattice structure with their corresponding energy spectrum of (a) SLG, (b) AA-BLG. (c) and (d) Yellow and black  bands correspond to electrons and holes in SLG while in AA-BLG they represent electron- and hole-like states. Red and blue bands represent the upper and lower Dirac cones in AA-BLG. (e) Schematic illustration of  delaminated BLG connected to AA-BLG, and (f) SLG attached to AA-BLG whose terminated edge of the top layer are either zigzag or armchair.          } \label{fig:Lattic_Spectrum}
\end{figure}

The crystalline structures of single layer graphene and AA-BLG are illustrated  in   Figs. \ref{fig:Lattic_Spectrum}(a, b) with the corresponding energy spectrum in Figs. \ref{fig:Lattic_Spectrum}(c, d), respectively.  SLG   has a hexagonal crystal structure  comprising two atoms $A$ and $B$ in its unit cell  with interatomic distance $a=0.142$ nm and intra-layer coupling $\gamma_0=3$ eV\cite{Zhang_2011}. In the AA-BLG the two SLG are placed exactly on top of each other    with a  direct inter-layer coupling $\gamma_1\approx0.2\ $eV \cite{Li2009,Xu_2010,Lobato_2011}, see dashed-green vertical lines in Fig. \ref{fig:Lattic_Spectrum}(d). Pristine AA-BLG has a linear energy spectrum that consists of two Dirac cones (lower and upper cones) shifted by $2\gamma_1$, see blue and red cones in Fig. \ref{fig:Lattic_Spectrum}(d). These two cones are completely decoupled\cite{Sanderson_2013} such that electron- and hole-like carriers are associated with each cone. 

The general form of the Hamiltonian within the continuum approximation  describing the charge carriers near the K-point in reciprocal space is a $4 \times 4$ matrix in the basis $\mathbf{\Psi}=(\Psi_{A 1},\Psi_{B 1},\Psi_{A 2},\Psi_{B 2})^{T}, $ whose elements refer to the sublattices in each layer. Transport in both  connected and disconnected regions can be described by the following Hamiltonian 
\begin{equation}\label{Hamiltonian}
\mathcal{H}=\left(
\begin{array}{cccc}
  v_{0} & v_{F}\hat{\pi}_{+} & \tau\gamma_{1} & 0 \\
  v_{F}\hat{\pi}_{-} & v_{0} &  0 & \tau\gamma_{1}\\
  \tau\gamma_{1} &  0 & v_{0}& v_{F}\hat{\pi}_{+} \\
  0 & \tau\gamma_{1}& v_{F}\hat{\pi}_{-} &v_{0} \\
\end{array}%
\right),
\end{equation}
where  $v_{F}\approx10^{6}$ m/s is the Fermi velocity\cite{Castro_Neto_2009}  of charge carriers in each graphene layer, $\hat\pi_{\pm}=p_{x}\pm ip_{y}$ denotes the momentum, $v_{0}$ is the strength of a local electrostatic potential. The coupling between the two graphene layers is controlled by the parameter $\tau $ through which we can ``\textit{switch on}" or ``\textit{switch off}" the inter-layer hopping between the sublattices. For $\tau =0$, the two layers are decoupled and the Hamiltonian reduces to two independent SLG sheets  while for AA-stacking we take $\tau =1$. The domain wall under consideration in this Article is, therefore, described by a local change in $\tau$  from zero to one. 

Finally, notice that for the second configuration of this study, where transport from a single layer into an AA-bilayer system is considered, the Hamiltonian in Eq. \eqref{Hamiltonian} does not suffice. Rather, one needs to resort to the $2\times 2$ upper-left block that describes transport in a single layer of graphene. The effect of the atomic structure on the electronic transport is in this case determined through the boundary conditions (BCs). 

The collimation occurs at the boundary between two stacking types. The terminated edge of AA-BLG can cross the lattice at any angle. At specific angles  there are  in general two distinct edges, namely, zigzag and armchair edges\cite{Nakanishi_2010}. Imposing  zigzag boundary can be established through two different ways, namely, ZZ1 and ZZ2 where the sublattices  $\phi_{B2}$ and $\phi_{A2}$ are set to be zero at the edge, respectively\cite{Nakanishi_2010}.  Note that the two types of  the zigzag edges are equivalent in AA-BLG such that $T_{ZZ1}(\phi)=T_{ZZ2}(-\phi)$, where $T$ is the transmission probability, while this is not the case for AB-BLG. This can be attributed to  the symmetric and asymmetric  inter-layer coupling in, respectively,  AA-BLG and AB-BLG. For the armchair edge, the single valley approximation is not valid anymore and thus the BCs are inter-valley mixed such that\cite{Nakanishi_2010}\begin{equation}
 \phi_{A2}^K-\phi_{A2}^{K^{'}}=0, \text{ and }
 \phi_{B2}^K+\phi_{B2}^{K^{'}}=0.
\end{equation}

\subsection{Semi-classical dynamics}\label{SCD}

To describe electron dynamics semi-classically one proceeds in two steps. We first use the quantum mechanical formalism to evaluate  transmission and reflection probabilities\cite{Abdullah2017,Barbier01_2010,Van_Duppen01_2013,Abdullah_2017}, and secondly determine the electron trajectories using the classical approach. Since the system is invariant along the $x-$direction, the  solution of the Schr\"odinger equation $\rm H{\bf{\Psi}}=E\bf{\Psi}$ can be written in a matrix form as  
\begin{equation}\label{Eq:Wavefun}
{\bf{\Psi}}(x,y)= M(y)C e^{ik_{x}x},
\end{equation}
where the  $4\times 4$ matrix $ M(y)$ represents the plane wave solutions, and the four-component vector $C$ contains the different coefficients expressing the relative weights of the  different traveling modes, which have to be set according to the propagating region. After obtaining the desired solutions on both sides of the domain wall, we then implement the transfer matrix together with appropriate boundary conditions to obtain the transmission and reflection probabilities.

\begin{figure}[t!]
\vspace{0.cm} 
\centering\graphicspath{{./Figures/}}
\includegraphics[width=1.65  in]{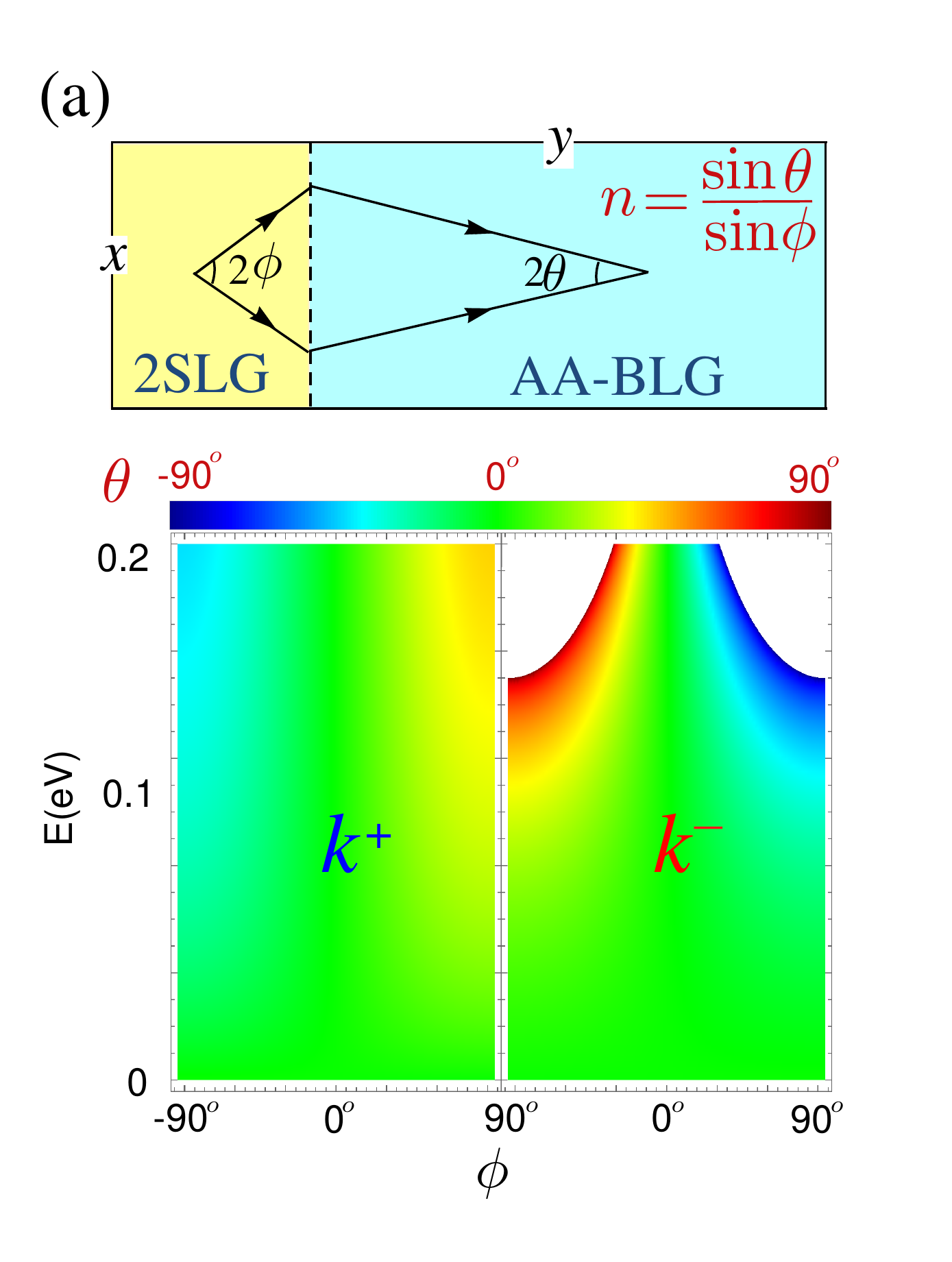}\
\includegraphics[width=1.65  in]{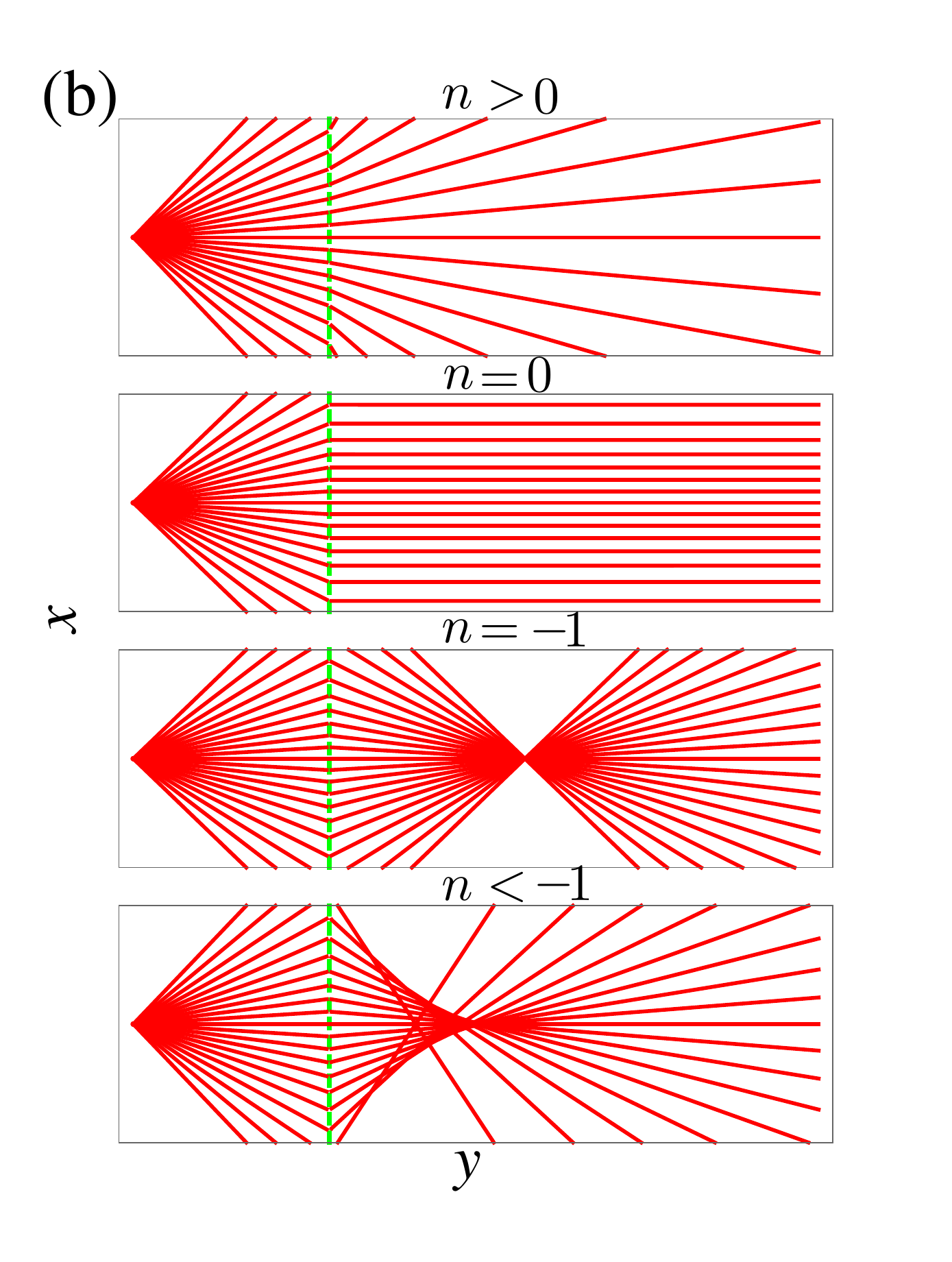}\\
\includegraphics[width=\linewidth]{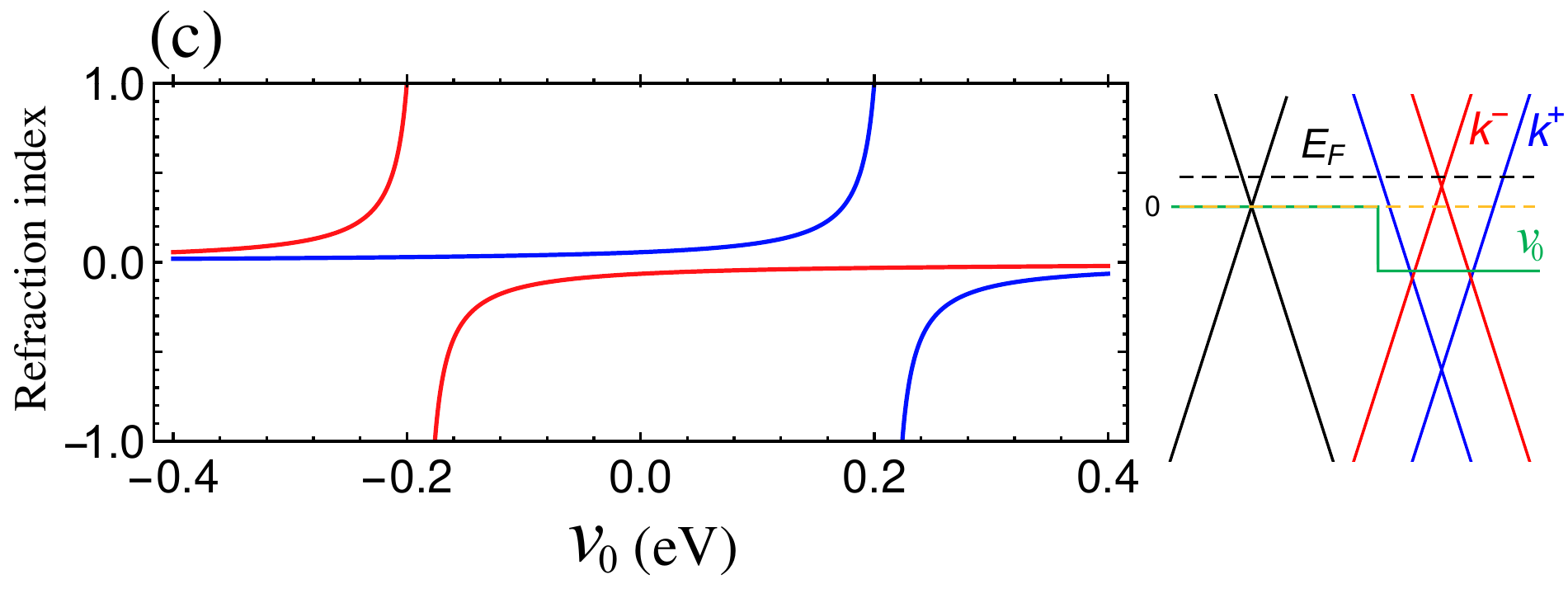}
\vspace{0.cm}
\caption{(Color online) (a) Top panel illustrates the 2SLG-BLG junction with  incident and transmitted electron beams  in the $x-y$ plane, while the bottom panel shows the transmitted angle $\theta$ as a function of the Fermi energy and the incident angle $\phi$ for  SLG-AA junctions with $v_0=0.1$ eV. (b) Classical trajectories of an electronic beam impinging on media  with different refraction indices.     (c)  Refraction index with the corresponding band diagram for    SLG-AA as a function of the electrostatic potential strength $v_0$ where the  Fermi energy   of the incident particles is $E=12$ meV. Blue and red curves correspond to the different modes in AA-BLG region.} \label{fig:ref_index}
\end{figure}

To calculate the electron trajectories, we assume a divergent beam starting from a focal point with a wave propagation given by the wave vector $\vec{k}$. The difference in wave vector between the connected and delaminated regions is determined by the relative refractive index\cite{Pereira2010,Cheianov2007,Milovanovic2015,Masirz2010,Barbier2010,Lee2015}
\begin{equation}\label{Eq:ref-ind}
n=\frac{\sin\theta}{\sin\phi}=\frac{k'_{y}}{k_y},
\end{equation}
 where $\phi$ and $\theta$ are the incident and transmitted angles, respectively, while $k'_y$ and $k_y$ are the wave vectors of the incident and transmitted electrons, respectively. For 2SLG-AA junction these wave vectors are given by
\begin{equation}\label{Eq:AB_Wave_vec}
k'_y=\frac{E}{v_F\hbar},\ \ k_y^{\pm}=\frac{1}{v_F\hbar}(\epsilon\pm\gamma_1),
\end{equation}
where $\epsilon=E-v_0$ and $E$ is the Fermi energy. Using the above equations one can obtain  the classical trajectories\cite{Reijnders2017,Park2011,Phong2016,Milovanovic2014,Peterfalvi2012}. In Fig. \ref{fig:ref_index}(a), we show the system geometry (top panel) and the   transmitted angle (bottom panel), according to Eq. \eqref{Eq:ref-ind}, associated with  the lower and upper cones. To achieve perfect collimation, the transmission angle must be zero which corresponds to zero  refraction index. The refraction index of electrons incident from SLG and transmitted into gated AA-BLG  is shown in Fig. \ref{fig:ref_index}(c) as a function of the electrostatic gate $v_0$. It is clear that the refraction index is almost zero in pristine AA-BLG, i.e. $v_0=0$. Henceforth, the gate will be considered zero and the calculations will be based only on    pristine AA-BLG. A schematic presentation of the  classical trajectories  of carriers with different refraction indices is shown in Fig. \ref{fig:ref_index}(b) and our interest is when $n=0$ where carriers move in one dimension.   

\begin{figure}[t!]
\vspace{0.cm}
\centering\graphicspath{{./Figures/}}
\includegraphics[width= \linewidth]{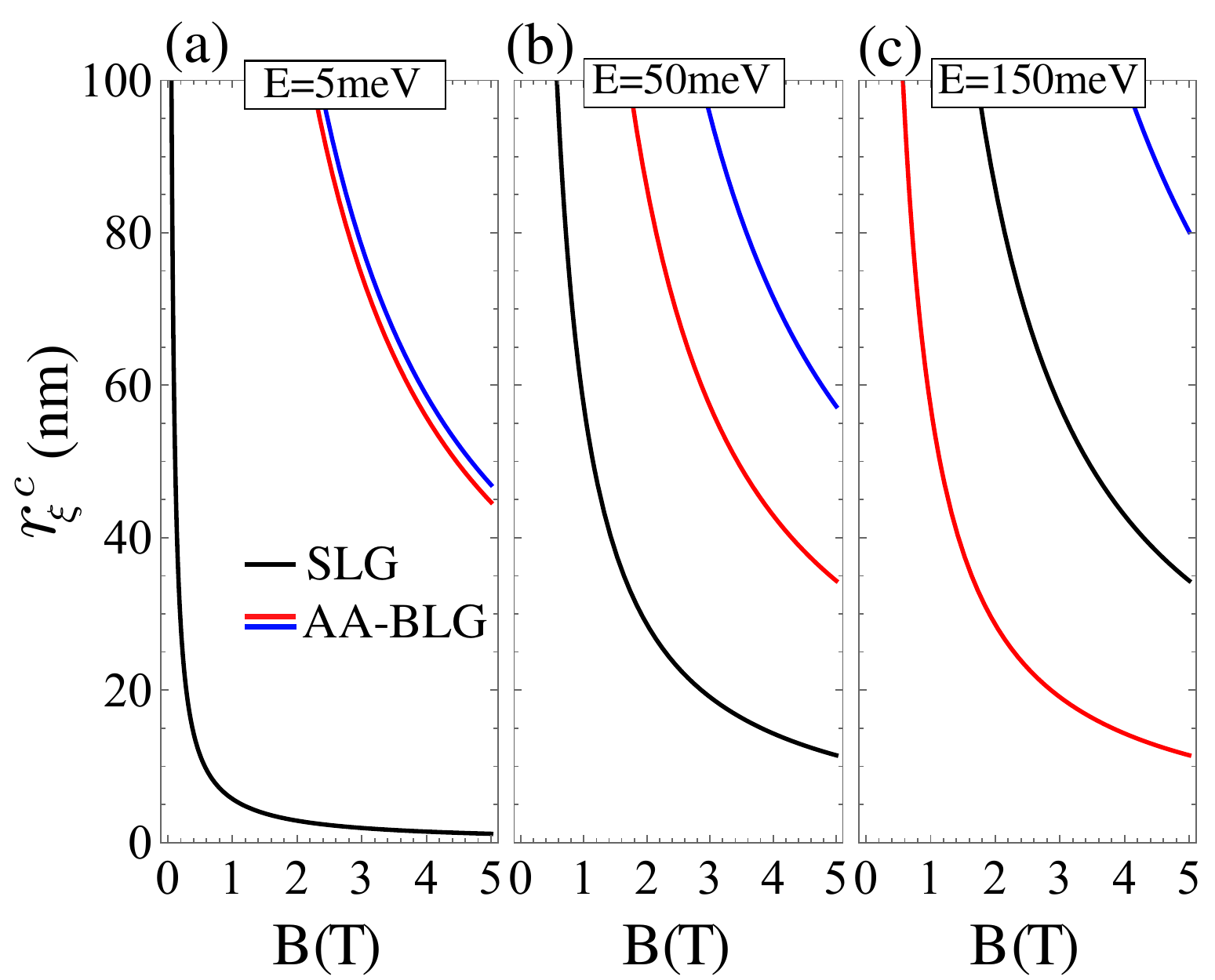}\
\vspace{0.cm}
\caption{(Color online) Cyclotron radius in pristine (black curves) SLG and in AA-BLG for different Fermi energy. The red and blue curves  correspond to the  upper and lower cones in AA-BLG, respectively.} \label{fig:cycl_radii}
\end{figure}

In the presence of a perpendicular magnetic field $\vec{B}$, the motion of the charge carriers follows a curved trajectory with curvature radius $r^{c}$.   In the ballistic transport regime, where the Fermi wavelength is much smaller than the geometric size of the system, the charge carriers can be treated as classical point-like particles. Thus, we can calculate the cyclotron radius  $r^{c}$ following Lorentz law described by
\begin{equation}\label{motion}
m\vec{a}=-e\vec{v}\times \vec{B},
\end{equation}
where $e$ is the elementary charge,   $\vec a$ and $\vec v$ are the  acceleration and speed of electron, respectively. Note that the electron's speed  $\left\vert \vec v \right\vert$  will be assumed here to be the Fermi speed $v_F$. The effective mass $m$ of a particle in an isotropic energy spectrum reads\cite{Ariel2013,Ashcroft1976,Zou2011}   \begin{equation}\label{energy}
m=\frac{\hbar^2}{2 \pi}\frac{dA(E)}{dE},
\end{equation}
where $A(E)$ indicates the area in $k-$ space enclosed by a constant energy contour $E$. This area is circular in single layer graphene and AA-BLG. Note that  depending on the energy curvature whether it is convex or concave, carriers can have a negative effective mass which is attributed to hole-like particles. Consequently, in the presence a magnetic field carriers with opposite sign of the effective mass will be deflected in opposite direction, as we will explore in the next sections. From Eqs.~\eqref{motion} and \eqref{energy} we can obtain the cyclotron radius for AA-BLG and SLG as follows:

\begin{equation}\label{radius_AA}
r_{\xi}^{c}(E)=\frac{\left\vert E\pm\xi\ \gamma_1 \right\vert }{ev_{F}\left\vert B \right\vert},
\end{equation} 
where $E$ is the Fermi energy and  $\xi=0$ or $1$ for SLG and AA-BLG, respectively. Finally, the  equations of motion in the $x-y$ plane can be written as
\begin{subequations}
\begin{equation}\label{x_motion}
x(t)=x_{\xi}-r_{\xi}^{c}\cos\left( \frac{v_{F}}{r_{\xi}^{c}} t+\Phi_{\xi} \right)+r_{\xi}^{c}\cos\left( \Phi_{\xi} \right),
\end{equation} 
\begin{equation}\label{y_motion}
y(t)=y_{\xi}+r_{\xi}^{c}\sin\left( \frac{v_{F}}{r_{\xi}^{c}} t+\Phi_{\xi} \right)-r_{\xi}^{c}\sin\left( \Phi_{\xi} \right),
\end{equation}
\end{subequations} 
where the point source of current is  at $(x_0,y_0)$  in the SLG region while $(x_1,y_1)$ indicates the point where the electron hits the domain wall. $\Phi_{0(1)}$ represents the incidence (transmission) angle    $\phi(\theta)$ described in the top panel of Fig. \ref{fig:ref_index}(a). Note that for $\xi=0$,  $ t$ is the time interval for the electron calculated once it is emitted from the current source  while  for $\xi=1$ it is the period of time calculated once  the electron enters the AA-BLG region. Using the above equations we can trace the trajectories of the charge carriers in a magnetic field. In Fig. \ref{fig:cycl_radii}, we show the cyclotron radii  for SLG and AA-BLG as a function of the magnetic  field at different Fermi energies. At low energy,  we see that the SLG cyclotron radius is sensitive to the magnetic field while in AA-BLG the cyclotron radii of the lower and upper cones (blue and red curves, respectively) are almost the same, see Fig. \ref{fig:cycl_radii}(a). Note that as a result of the spectrum resemblance of SLG and AA-BLG, we have  $r^{c}_{AA}(E)=r^{c}_{SL}(E\pm\gamma_1)$ which  can be inferred from Figs. \ref{fig:cycl_radii}(b, c). 

\subsection{Wave packet dynamics}\label{Sec:WD}

To calculate the quantum electronic trajectories using a wave packet, we apply the nearest-neighbor tight-binding model Hamiltonian for the description of electrons in a BLG system associated with the split-operator technique\cite{Batista2018,Costa2017,Chaves2015a,Cavalcante2016,Costa2012,da_Costa_2015,Chaves2010,Chaves2009,Degani2010,Chaves2015,Rakhimov2011}. We have added to this technique the van der Waals domain walls as a local variation in the inter-layer coupling parameter as described by the parameter $\tau$ in Eq. \eqref{Hamiltonian}. Following the numerical procedure developed in details by da Costa et al. in Ref.~[\onlinecite{da_Costa_2015}], that is based on the split-operator technique, we calculate the time-evolution of the wave packet for two different set-ups composed of two disconnected SLG bounded with a AA-stacked BLG and one SLG bounded with a AA-stacked BLG. 

Among the many different techniques to treat the formal solution of the time-evolution problem, such as Green's functions techniques\cite{Kramer2011}, here we decided to choose the split-operator technique, since using this approach, one has the possibility of observing the transmitted and reflected trajectories of the total wave packet describing the electron propagating through the system, as well as the separated trajectories in each layer and also the scattered trajectories projected on the different Dirac cones. Moreover, this approach has the advantage of being faster and easier than  e.g.~Green's functions techniques and, is pedagogical and physically a transparent approach for the understanding of transport properties in quantum systems, like the ones studied here.

The wave packet propagates in a system obeying the time-dependent Schr\"odinger equation $i \hbar \partial \mathbf{\Psi}(\vec{r},t) = H \mathbf{\Psi}(\vec{r},t)$, where the Hamiltonian $H$ is the nearest-neighbor tight-binding Hamiltonian given by
\begin{equation}\label{TBHamiltonian}
H_{TB}= \sum_{i\neq j}(\gamma_{ij} c_i^{\dagger}c_j + h.c) +  \sum_{i}(\epsilon_i+V_i)c^{\dagger}_ic_i,
\end{equation}
where $c_{i}$($c_{i}^{\dagger}$) annihilates (creates) an electron in site $i$ with on-site energy $\epsilon_i$, $\gamma_{ij}$ is the nearest-hopping energy between adjacent atoms $i$ and $j$, and $V_i$ is the on-site potential. The effect of an external magnetic field can be introduced in the tight-binding model by including a phase in the interlayer hopping parameters according to the Peierls substitution $\gamma_{ij} \rightarrow \gamma_{ij}\exp\left[i\frac{e}{\hbar}\int_j^i\vec{A} \cdot d\vec{l}\right]$, where $\vec{A}$ is the vector potential describing the magnetic field. We conveniently choose the Landau gauge $\vec{A} = (0,Bx,0)$, giving a magnetic field $\vec{B} = B\hat{z}$. The BLG flake considered in our tight-binding calculations has $3601 \times 1000$ atoms in each layer, thus being like a rectangle with dimensions of $\approx 213 \times 443$ nm$^2$. Such a large ribbon-like flake is necessary, in order to avoid edge scattering by the wave packet. Therefore, no absorption potential at the boundaries is needed to avoid spurious reflection.
\begin{figure}[t!]
\vspace{0.cm}
\centering\graphicspath{{./Figures/}}
\includegraphics[width=1.5  in]{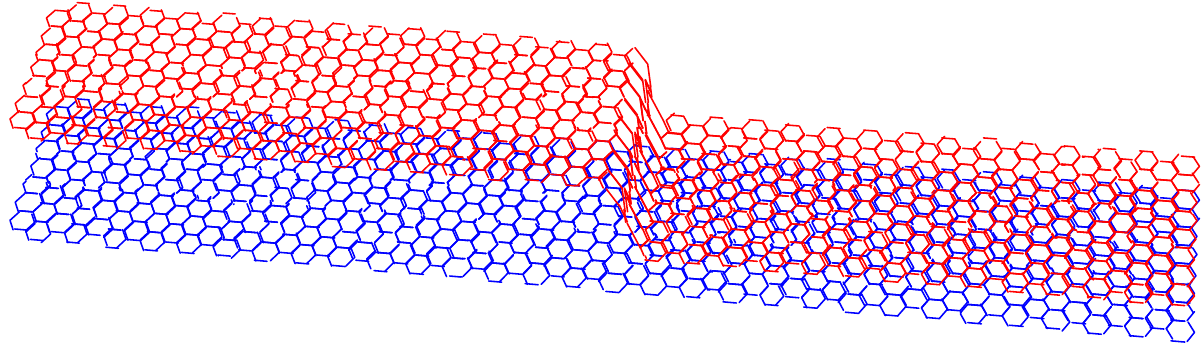}\\
\includegraphics[width= \linewidth ]{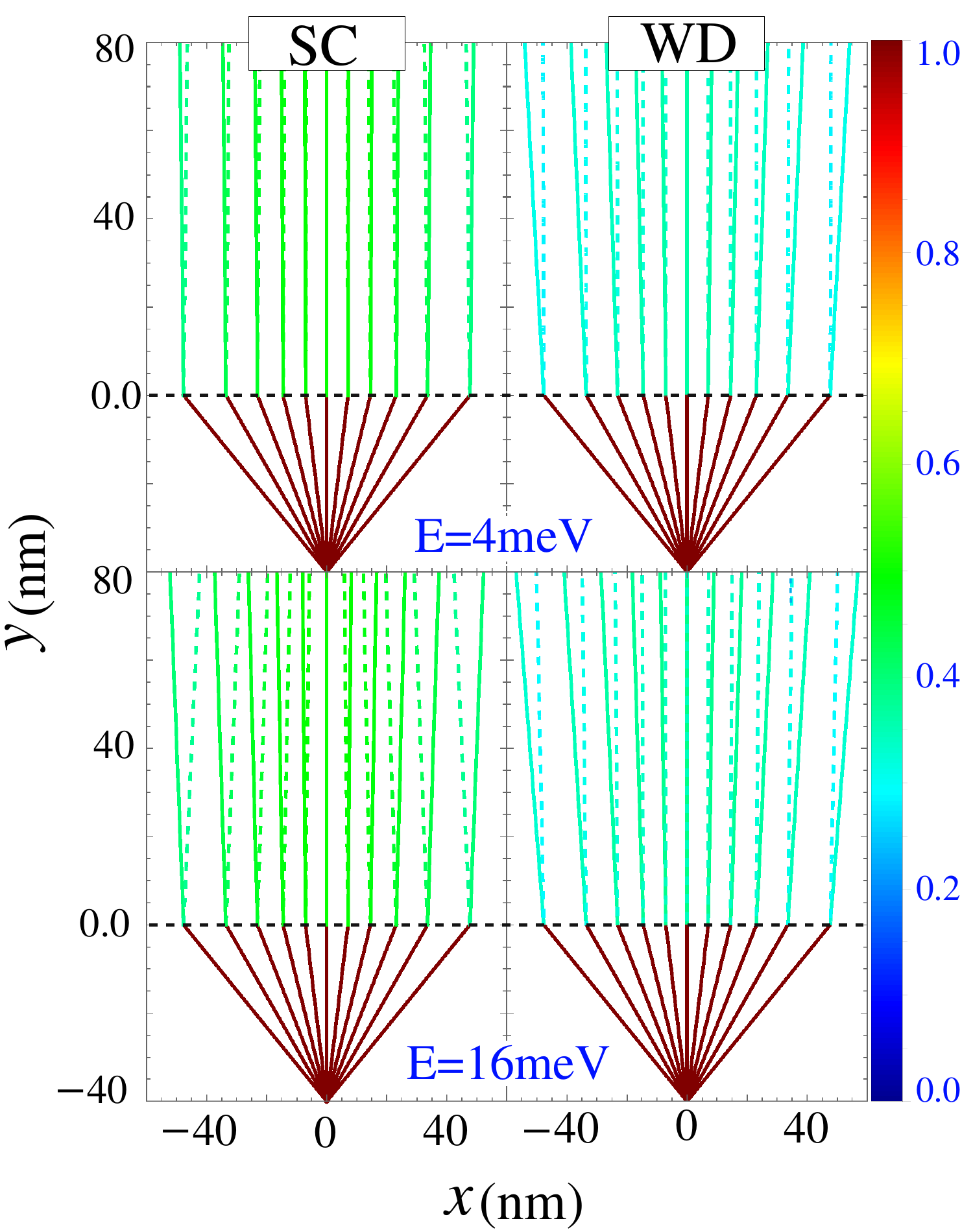}\\
\vspace{0.cm}
\caption{(Color online) Scattering from 2SLG  into lower $T_+$ (solid lines) and upper $T_-$ (dashed lines) cones in AA-BLG with different incident energies. Both 2SLG and AA-BLG are pristine where left and right columns show trajectories obtained from SC and WD approaches, respectively. Color bar represents the transmission probability where $R+T=1$ and $T=T_++T_-$.  } \label{fig:Traj_2SLG_AA}
\end{figure}
\begin{figure}[t!]
\vspace{0.cm}
\centering\graphicspath{{./Figures/}}
\includegraphics[width=1.5  in]{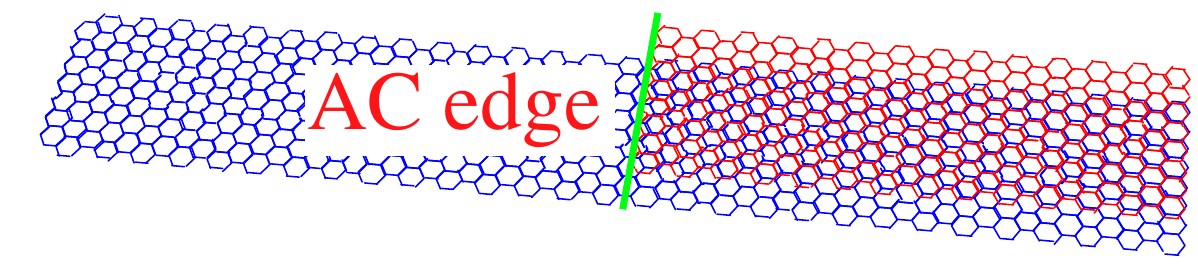}\\
\includegraphics[width= \linewidth]{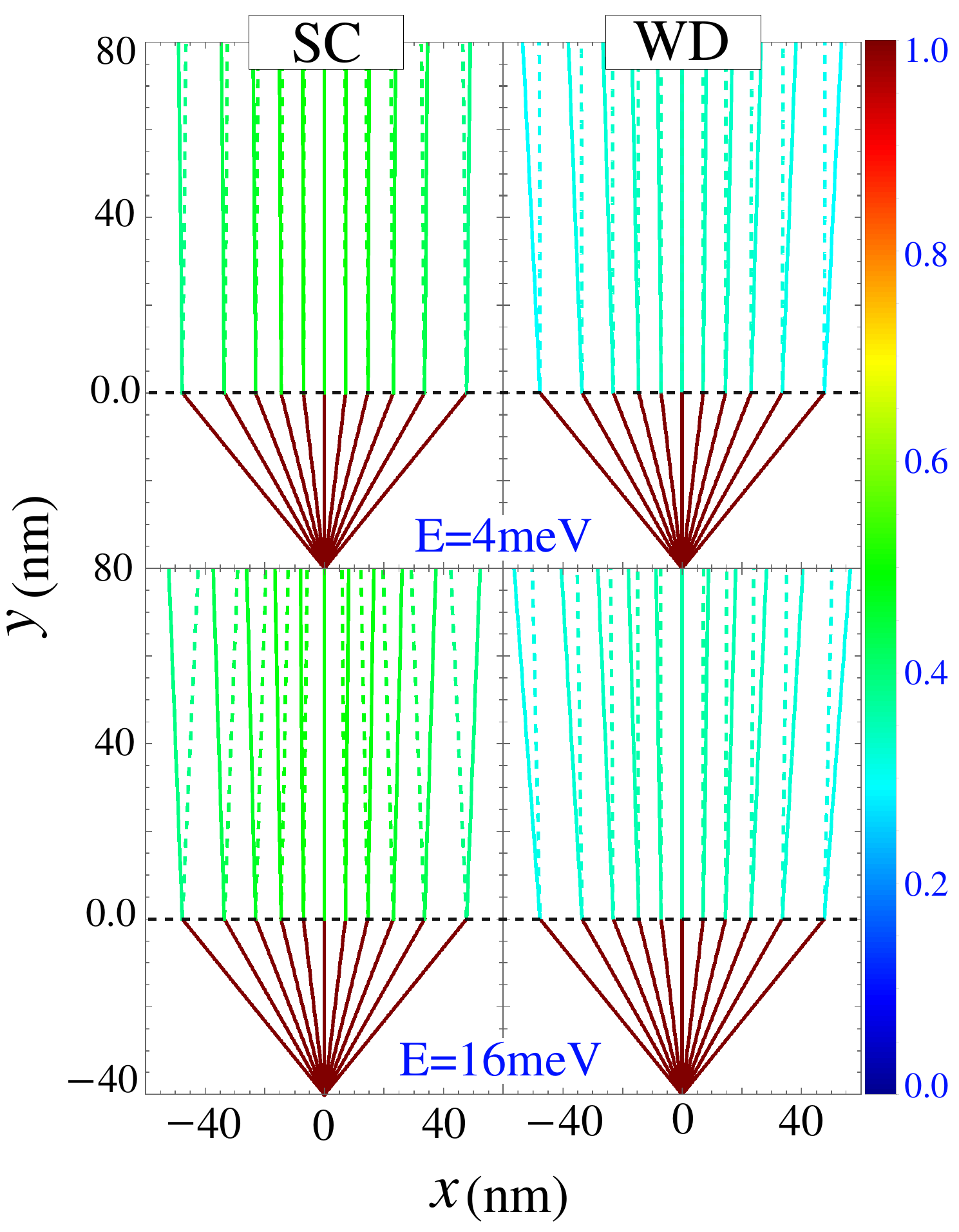}\\
\vspace{0.cm}
\caption{(Color online) The same as in Fig. \ref{fig:Traj_2SLG_AA}, but the scattering here is from SLG to AA-BLG, whose top layer possesses armchair edge at the interface.} \label{fig:Traj_AC_AA}
\end{figure}

The initial wave packet is taken as a circularly symmetric Gaussian distribution, multiplied by a four spinor in atomic orbital basis  $[\psi_{A1}, \psi_{B1}, \psi_{A2}, \psi_{B2}]^T$ and by a plane wave with wave vector $\vec{k} = (k_x,k_y)$, which gives the wave packet a non-zero average momentum, defined as
\begin{align}
{\bf\Psi}(\vec{r},t\hspace{-0.1cm}=\hspace{-0.1cm}0) \hspace{-0.1cm}=\hspace{-0.1cm} N\hspace{-0.1cm}
\left(\hspace{-0.1cm}\begin{array}{c}
\psi_{A1}\\
\psi_{B1}\\
\psi_{A2}\\
\psi_{B2}
\end{array}\hspace{-0.1cm}\right)\exp\hspace{-0.1cm}\left[\hspace{-0.1cm}{-\frac{(x\hspace{-0.1cm}-\hspace{-0.1cm}x_0)^2\hspace{-0.1cm}+\hspace{-0.1cm}(y\hspace{-0.1cm}-\hspace{-0.1cm}y_0)^2}{2d^2}\hspace{-0.1cm}+\hspace{-0.1cm}i\vec{k}\cdot\vec{r}}\right]
\label{eq:start_fourspinor}
,\end{align}
where $N$ is a normalization factor, $(x_0,y_0)$ are the coordinates of the initial position of center of the Gaussian wave packet, and $d$ is its width. For all studied cases, the width of the Gaussian wave packet was taken as $d = 10$ nm and its initial position as $(x_0,y_0) = (0, - 40)$ nm. 

The propagation direction is determined by the pseudospin polarization of the wave packet and plays an important role in defining the direction of propagation. It is characterized by the pseudospin polarization angle $\Theta$, such as $\left(1 \mbox{~,~} e^{i\Theta}\right)^T$ for the components in each layer. The choice of the angle $\Theta$ depends also on which energy valley the initial wave packet is situated\cite{Chaves2015,Rakhimov2011,Chaves2010,da_Costa_2015,Costa2012,Costa2017}. Our choice for the propagation direction here is based on the  knowledge reported in  literature for wave packet time evolution on monolayer\cite{Rakhimov2011,Chaves2010,Costa2012} and bilayer\cite{da_Costa_2015} graphene systems and the consequences of the \textit{Zitterbewegung} effect on the wave packet trajectory\cite{Maksimova2008}. We assume the $y$-direction as the preferential propagation direction, since the average position of  electronic motion along this direction is  less affected by the oscillatory behavior caused by the Zitterbewegung\cite{da_Costa_2015}.
\begin{figure}[t!]
\vspace{0.cm}
\centering\graphicspath{{./Figures/}}
\includegraphics[width= \linewidth  ]{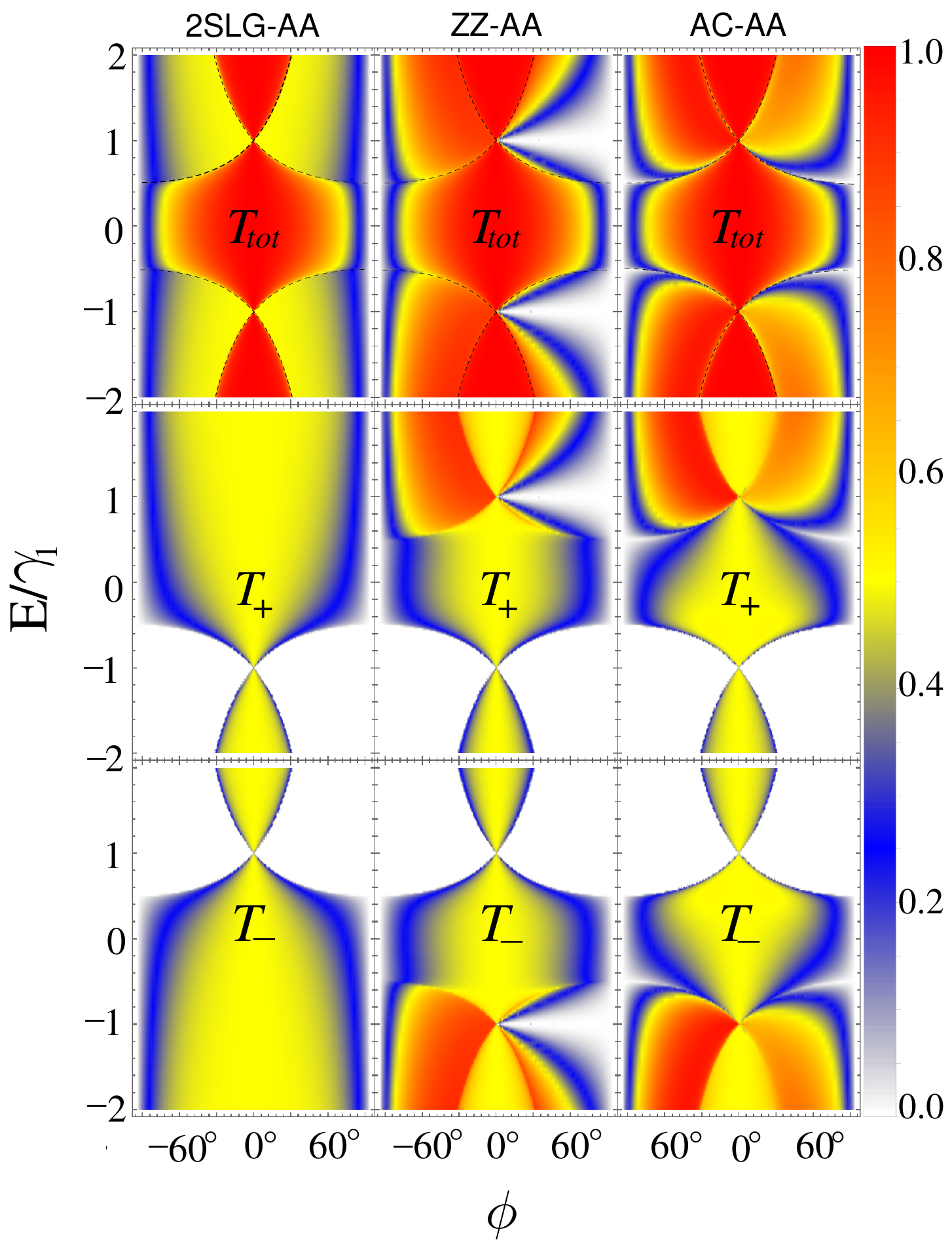}\
\vspace{0.cm}
\caption{(Color online) Comparison between the transmission probabilities obtained from the SC approach for 2SLG-AA and SLG-AA with zigzag- and armchair-edges with  $v_0=0$.  $T_+$ and $T_-$ are the cone transmission probabilities where carriers scatter into the lower and upper cones, respectively.  } \label{SC_Transmission}
\end{figure}
\begin{figure}[t!]
\vspace{0.cm}
\centering\graphicspath{{./Figures/}}
\includegraphics[width= \linewidth  ]{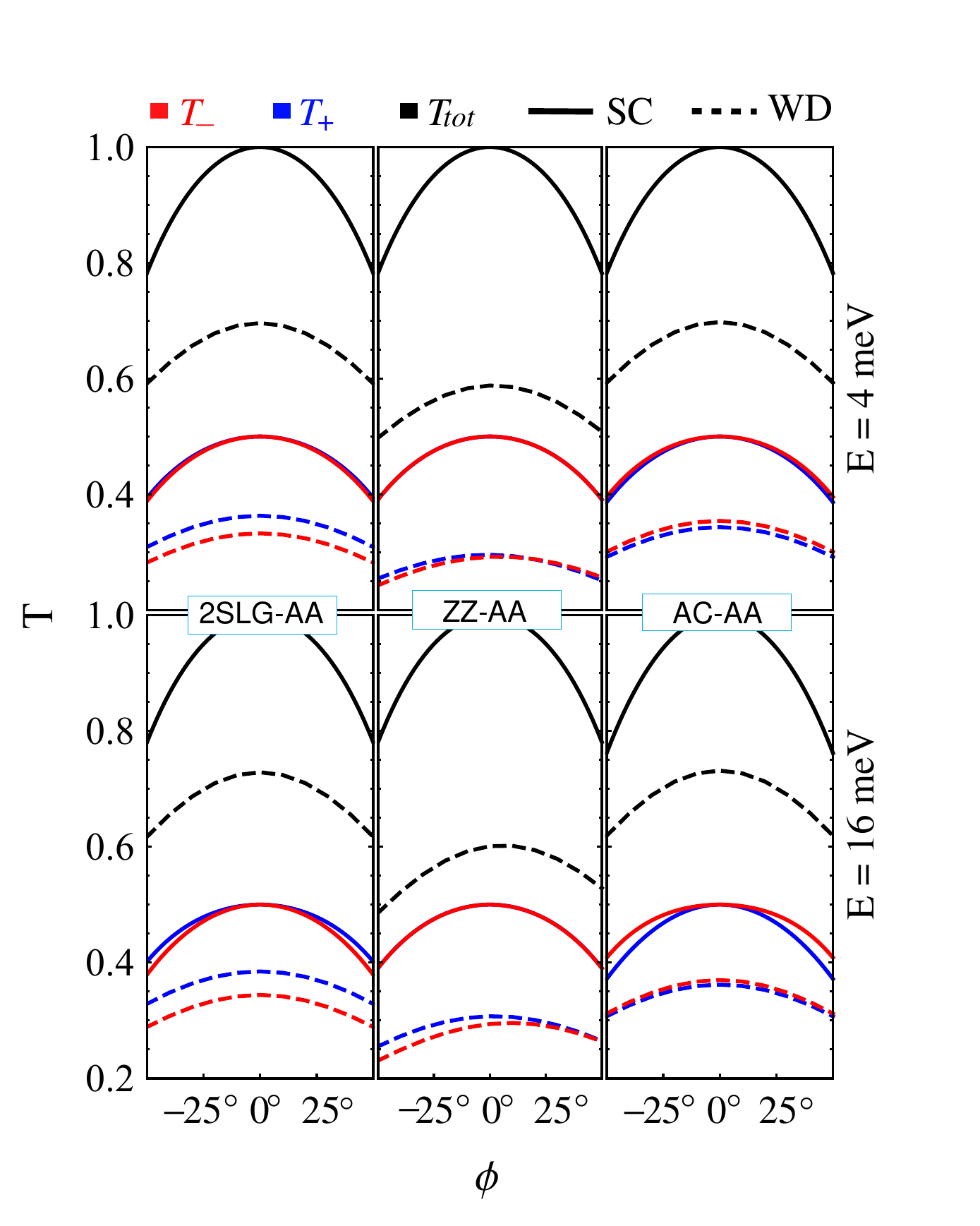}\
\vspace{0.cm}
\caption{(Color online) Comparison between the transmission probabilities obtained from the wave packet dynamics (WD) and semi-classical approach (SC). The incident energies are $4$ and $16$ meV for top and bottom rows, respectively, while the electrostatic potentialis  $v_0=0$.} \label{fig:Trans_comparison}
\end{figure}

The initial wave vector is taken in the vicinity of the Dirac point $\vec{k} = (k_x, k_y) + \vec{K}$, where $\vec{K} = (0, \pm 4\pi/(3\sqrt{3a}))$ represents the two non-equivalent $K$ and $K'$ points. As we intend to investigate the wave packet trajectories for different propagation angles and their probabilities, we run the simulation for each system configuration, such as e.g. initial propagation angle, initial wave vector and energy, and then as the Gaussian wave packet propagates, we calculate for each time step the amount of transmission ($T(t)$) and reflection ($R(t)$) and find the electron after $(y>0)$ and before $(y<0)$ the interface at $y=0$, respectively, as the integral of the square modulus of the normalized wave packet in that region, given by
\begin{subequations}
\begin{align}
T(t) &=\int_{-\infty}^{\infty} dx \int_{0}^{\infty} dy |\mathbf{\Psi}(x,y,t)|^2, \\ 
R(t) &=\int_{-\infty}^{\infty} dx \int_{-\infty}^{0} dy |\mathbf{\Psi}(x,y,t)|^2,
\end{align}
\end{subequations}
and the total average position, i.e. the trajectory of the center of mass $\langle \vec{r}\rangle$ of the wave packet, that is calculated for each time step by computing
\begin{subequations}
\begin{align}
\langle x(t)\rangle \hspace{-0.1cm}&=\hspace{-0.125cm}\int_{-\infty}^{\infty} \hspace{-0.125cm}dx\hspace{-0.125cm} \int_{-\infty}^{\infty} \hspace{-0.125cm}dy  |\Psi(x,y,t)|^2x, \\ 
\langle y(t)\rangle \hspace{-0.1cm}&=\hspace{-0.125cm}\int_{-\infty}^{\infty} \hspace{-0.125cm}dx\hspace{-0.125cm} \int_{-\infty}^{\infty} \hspace{-0.125cm}dy  |\Psi(x,y,t)|^2 y.
\end{align}
\end{subequations}
For larger $t$, the value of the transmission (reflection) probability integral increases (decreases) with time until it converges to a number. This number is then considered to be the transmission (reflection) probability of such system configuration, i.e. $T = T(t\rightarrow\infty)$ ($R = R(t\rightarrow\infty)$).
\begin{figure}[t!]
\vspace{0.cm}
\centering\graphicspath{{./Figures/}}
\includegraphics[width= \linewidth  ]{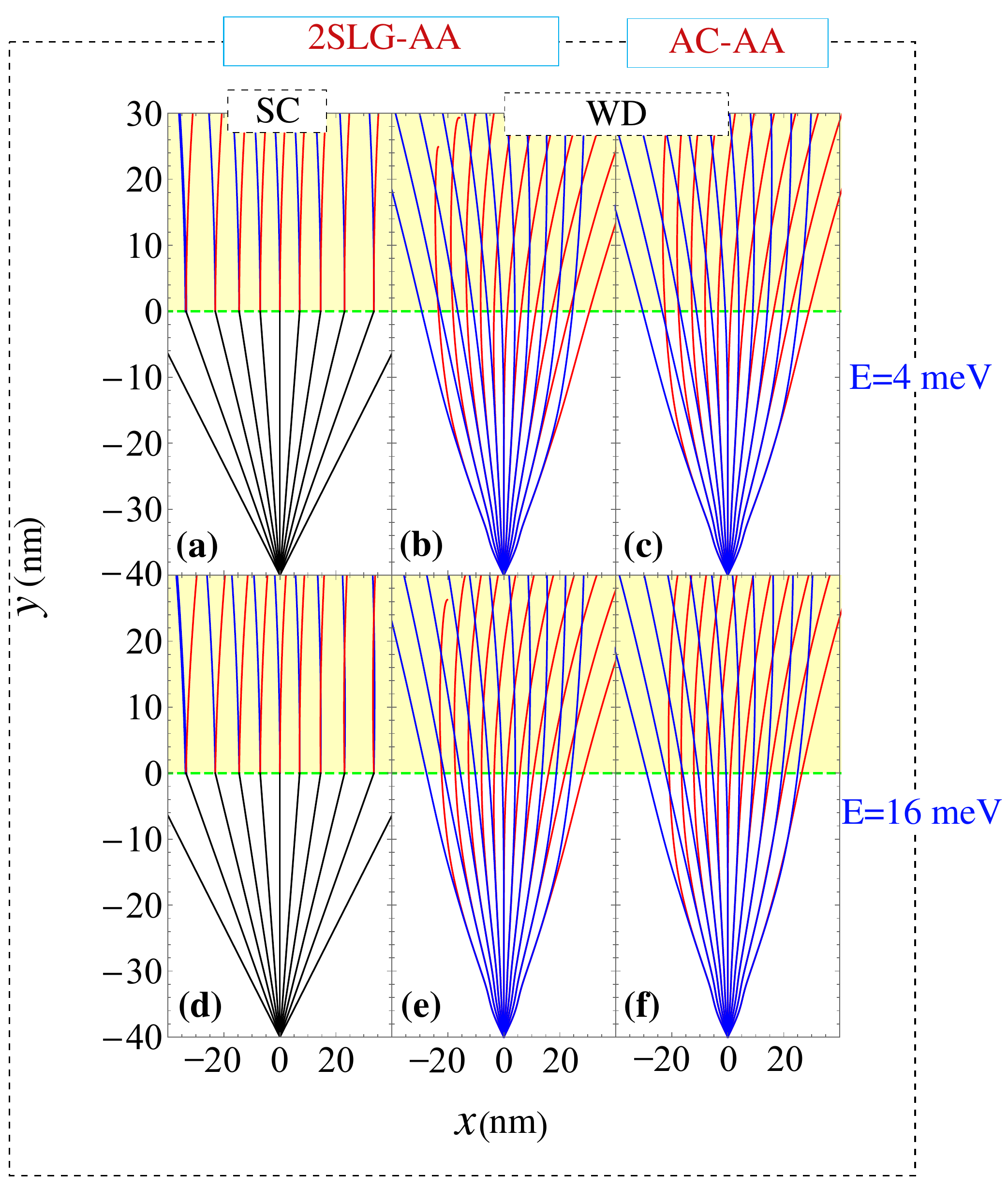}\

\vspace{0.cm}
\caption{(Color online) SC and WD classical trajectories of the charge carriers scattering from 2SLG into AA-BLG  and from SLG into AA-BLG with AC-edge in the presence of a perpendicular  magnetic field $B=1$T (only in the yellow region $y$\textgreater40 nm) in the $x-y$ plane  with   $v_0=0$ and at different Fermi energies.  SC gives the same results for both systems (left column), while WD provides  a slight difference for 2SLG-AA and AC-AA (middle and right columns). Red and blue trajectories correspond to scattering into the upper $k^-$ and lower $k^+$ cones in AA-BLG as indicated in the top of Fig. \ref{fig:ref_index}(c).} \label{fig:Class_Tranj_B}
\end{figure}
Note that, essentially, a wave packet is actually a linear combination of plane-waves, where the wave packet width represents a distribution of momenta and, consequently, of energy. In this sense, we are investigating the dynamics of a distribution of plane-waves with different energies around some average value, whose width can be even related e.g.~to the temperature of the system. A large wave packet in real space implies a narrow wave packet in $k$-space, thus it will be composed of a distribution of plane-waves with different velocities and, therefore, exhibits a strong decay in time. We have checked that the wave packet width in real space considered in our calculations is appropriate for the proposed problem, being large enough to avoid significant changes of the wave packet within the time scale of interest.
\begin{figure*}[t!]
\centerline\centering\graphicspath{{./Figures/}}
\includegraphics[width = \linewidth]{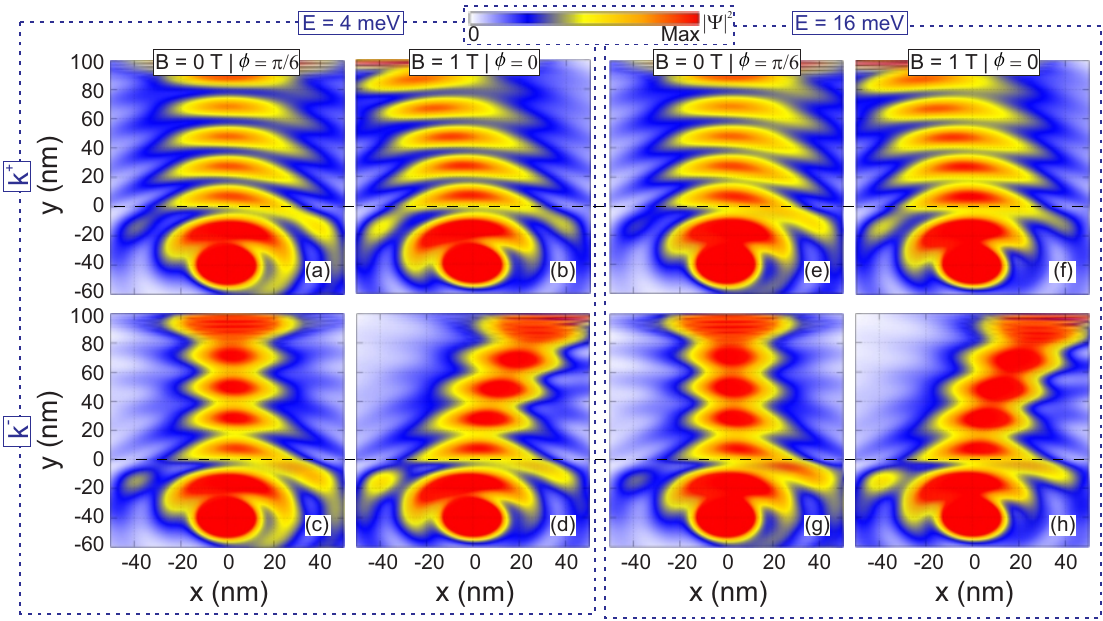}
\caption{(Color online) Contour plots of the time evolution for the squared modulus of the Gaussian wave function scattering from 2SLG into AA-stacked BLG with an initial energy (a-b) $E = 4$ meV and (e-h) $E = 16$ meV, for an incident angle $\phi=\pi/6$. The   magnetic field was assumed to be (a, c, e, g) $B=0$  and (b, d, f, h) $B=1$ T.  Solid-dashed black line indicates the interface of the junction. Top (bottom) panels correspond to the lower $k^+$ (upper $k^-$) cones in the AA-BLG spectrum.} \label{fig:psitracker.AA}
\end{figure*}

As mentioned before, the propagation of charge carriers in AA-BLG can be described as belonging to the upper or lower cone, respectively denoted by red and blue in Fig. \ref{fig:Lattic_Spectrum}(d). In order to investigate the wave packet scattering to these upper and lower Dirac cones $k^+$ and $k^-$, one can find  a unitary transformation $U$  that block-diagonalizes our Hamiltonian in Eq. \eqref{Hamiltonian}, such transformation reads 
\begin{equation}\label{eq.U}
U = \frac{1}{\sqrt{2}}\left(\begin{tabular}{cccc}
$1$ & $0$ & $1$ & $0$\\
$0$ & $1$ & $0$ & $1$\\
$1$ & $0$ & $-1$ & $0$\\
$0$ & $1$ & $0$ & $-1$\end{tabular} \right).
\end{equation}
Applying this to the four-spinor in Eq. \eqref{eq:start_fourspinor} forms symmetric and antisymmetric combinations of the top and bottom layer wave functions components, i.e. 
\begin{equation}
U\mathbf{\Psi} = \mathbf{\Psi}' = \frac{1}{\sqrt{2}}\left(\begin{tabular}{ccc} 
$\psi_{A2}+\psi_{A1}$\\
$\psi_{B2}+\psi_{B1}$\\
$\psi_{A2}-\psi_{A1}$\\
$\psi_{B2}-\psi_{B1}$ \end{tabular} \right).
\end{equation} 
The symmetric and antisymmetric components correspond to the $k^+$ and $k^-$ energy bands (For more details see Refs.~[\onlinecite{Abdullah2017}]). In our results for AA-BLG case, we use the above wave function to calculate the center mass position and the probability amplitudes. 

\section{Results and discussions} \label{Results}
\subsection{Without magnetic field}
Before we proceed to show the electron collimation in different  systems, we would like to remind the reader of the following:
there are three different junctions under  consideration, namely, 2SLG-AA, ZZ-AA, and AC-AA.
For the SC, the classical trajectories in the three configurations are the same since they  depend only on the radius of the Fermi circle on both sides of the junction. However, the transmission probability associated with each system is indeed different. On the other hand, for WD the electron trajectories and transmission probability are distinct in 2SLG-AA and AC-AA, while the results for ZZ-AA are strongly obscured due to the strong \textit{Zitterbewegung} effect along the zigzag edge as discussed in Sec. \ref{Sec:WD}. 

Additionally, the fact that the  lower and upper cones in AA-BLG are decoupled means that each cone exhibits electron- and hole-like carriers. For example, for $\gamma_1>E>0$ electron- and hole-like carriers emerge from the lower and upper cones, respectively. Consequently, there will be two different types of  collimated beams coming from the two cones as will henceforth be seen. 

In Fig. \ref{fig:Traj_2SLG_AA} we show the carrier collimation through a domain wall that separates 2SLG and AA-BLG obtained from both SC and WD calculations with different Fermi energies. The point source is positioned at $(x,y)=(0,y_0)$ nm and  electrons impinge  on the domain wall located at the origin ($y=0$), afterwards they scatter to either lower (solid) or upper (dashed) cones with different transmission angles. Both approaches show a strong agreement for carrier trajectories. For example, according to SC the refraction index   associated with the upper cone for   $E=4$ meV is $n=-0.0041$  while the   WD calculations give $n=-0.0039$. The plus and minus signs of  the refraction index  reveal that the respective charge carriers will diverge and converge, respectively, at  large distance. The transmission  probabilities obtained from the two approaches  agreed qualitatively as will be explained later. Experimentally, it is often found that some islands in a sample have single layer graphene  connected to bilayer graphene flakes\cite{Yan2016,Clark_2014}. In Fig. \ref{fig:Traj_AC_AA}, we show the carrier trajectories through  such structure. We notice that  even though the  transmission probabilities are slightly altered, the system still attains  collimation. We can say that the results are almost identical for 2SLG-AA and AC-AA as depicted in Figs. (\ref{fig:Traj_2SLG_AA}, \ref{fig:Traj_AC_AA}), respectively. 

To validate this understanding and quantitatively determine the degree of agreement, we next carry out a transmission comparison between different systems and approaches.
Using the SC approach, we show in Fig. \ref{SC_Transmission} the cone as well as the total transmission probabilities in 2SLG-AA, ZZ-AA, and AC-AA systems. In the cone channels $T_+$ and $T_-$, the charge carriers scatter from SLG region into the lower and upper cones, respectively. In 2SLG-AA system, the transmission is symmetric with respect to  normal incidence, while it becomes asymmetric   in ZZ-AA and AC-AA systems at high energy. Such an asymmetry feature is a manifestation of  breaking  the  inversion symmetry  in the system. Notice that the transmission remains symmetric in the $E-\phi$ regions  where both modes $k^+$ and $k^-$ are propagating and the asymmetry feature only appears when one of them becomes evanescent. The critical energy that separates these two domains are given by $E_{c}^{\pm}(\phi)=\pm\gamma_1/(1+\sin\phi)$  and is superimposed as dashed-black curves on  $T_{tot}$ in Fig. \ref{SC_Transmission}. The critical energy decreases with increasing  incident angle which reaches $E_c^{\pm}=\pm\gamma_1/2$  for $\phi=\pi/2$. Therefore, within this energy range, the electron beam is symmetrically collimated. Moreover,  within the same energy range the intensity of the collimated beam is almost the   same  for all systems.
Note that in the other valley $K'$ the total transmission probability in ZZ-AA and AC-AA attains the following   symmetry  $T_K(\phi)=T_{K'}(-\phi)$.
Note that if the edge crosses the lattice at arbitrary angle then such edge would be a mixed edge such that it is locally posses ZZ and AC boundaries. Since the transmission probabilities of both types are almost the same for low energies, we can safely assume that the mixed edge will not significantly alter the transmission probability and thus collimation is maintained since since the radius of Fermi circles in both sides of the junction remains unchanged regardless the edge type.      

For comparison with the WD calculations, we show in Fig. \ref{fig:Trans_comparison} the  transmission probabilities as a function of the incident angle at two different energies. The fundamental   characteristics   of the system are qualitatively  captured by both approaches. Of particular importance is the deviation  in the cone transmission at higher incident angles in 2SLG-AA and AC-AA. At normal incidence and in the SC picture, the cone channels are  equal, such that  $T_+=T_-=1/2$, while for oblique angles they start deviating from each other. For 2SLG-AA junction, we notice  that $T_+>T_-$ while it is reversed for AC-AA as can be inferred from the solid blue and red curves in Fig. \ref{fig:Trans_comparison}. This behaviour is also captured by the WD as can be seen from the dashed blue and red curves.  For ZZ-AA, the WD results for the transmission profile is asymmetric with respect to normal incidence. The reason for this is that the energy tail of the wave packet reaches the region where  one of the modes is evanescent as we explained earlier. Furthermore, it is clear that transmission amplitudes from  SC and WD do  not   match precisely. For example, at normal incidence $T_{\rm{tot}}$ is always unity for all systems according to SC, while it is significantly reduced in WD. The reason for this  difference is due to the fact that we consider a plane wave in SC approach with single  energy and momentum value. In contrasts WD uses a  wave packet that defines a  burst of particles with  a    momenta distribution $\hbar\Delta k_x$. Thus a perfect transmission is not expected since only part of the  wave packet coincides with normal incidence which will be completely transmitted.  While the part  associated with $k_x\neq0$   will be partially transmitted and reflected\cite{Rakhimov2011}. 
\subsection{With magnetic field}

So far, we have shown the electron collimation through different configurations in the absence of a magnetic field. Gaining  control over  the direction of the electron beams       can be realized  through a magnetic field without losing collimation.   To  examine the effect of the magnetic field on the collimated beams,  we assume that the magnetic field is applied only in AA-BLG region, i.e. for $y>0$. This can be justified by considering that the electron point source is located near the domain wall such that the distance is much smaller than $r^c_{SL}$. Note that even if a global magnetic field is subject to the system, the directional collimation will be maintained as long as $r^c_{SL}\gg \left\vert y_0 \right\vert $. To assess the effect of the magnetic field, we calculate the classical trajectories in 2SLG-AA and AC-AA using SC and WD as shown in Fig. \ref{fig:Class_Tranj_B}. We consider an electron beams with maximum incidence angles $\phi=\pm50^o$. The  essence of SC approach lies in expressing  the relative refraction index $n$ in terms of the wave vectors on both sides of the domain wall. Consequently, the classical trajectories for all considered configurations in the current paper are the same; thus, we show in Fig.  \ref{fig:Class_Tranj_B}.   the trajectories for only 2SLG-AA. This is also confirmed by the WD calculation where it shows that the trajectories for 2SLG-AA and AC-AA are almost the same, see Figs.  \ref{fig:Class_Tranj_B}(b,c) and (e,f).  Both SC and WD show contributions from two types of trajectories which is a direct consequence of the electron- and hole-like nature of the carriers associated with the lower and upper cones, respectively. The two trajectories are steered by the magnetic field in diametrically opposite directions.  

Finally, to clearly visualize the effect of the magnetic field on the whole wave packet, we show in  Fig. \ref{fig:psitracker.AA} the contour plots of the time evolution  for the squared modulus of the Gaussian wave for 2SLG-AA. We set the incidence angle  to be $\phi=0$ and $\phi=30^o$ and show the scattering to each cone separately in the presence and absence of  magnetic field, respectively. For $B=0$, once the wave packet reaches the domain wall it starts moving  nearly along the $y-$direction, see Figs.  \ref{fig:psitracker.AA}(a, c) and (e, g) and compare with the trajectories in Fig. \ref{fig:Traj_2SLG_AA}. In the presence of a magnetic field, the wave packets corresponding to lower and upper cones are steered in different directions without losing  collimation. Note that due its  spatial spread  the  wave packet feels  the magnetic field  before  its  center reaches the interface and this is  clearly  seen in Fig. \ref{fig:Class_Tranj_B}. Such behaviour is a manifestation of the quantum non-locality nature of  the charge carriers in graphene. 

It is important to point up that within the tight-binding model, the effects like the Landau levels in the presence of a perpendicular magnetic field are already embedded in the model, such that we do not need to take nothing more in consideration to take this issue into account, as well as, regardless of the value of the chosen magnetic field, the tight-binding model in the WD simulation takes into account all the consequences of its inclusion. Therefore, for convenience we chose such magnetic field values in order to consider a slightly smaller BLG sample, since it can become computationally expensive for larger structures, keeping in mind that enlarging the sample by a factor $\beta$ will result in a similar collimation effect when reducing the magnetic field by a factor $1/\beta$.

\section{conclusion}\label{Concl}
In conclusion, we have studied electron scattering through locally delaminated AA-BLG systems with two different domain walls. Within the mesoscopic limit where electron current is well approximated by classical trajectories, we presented the SC model that combines quantum mechanical calculations of the transmission probabilities with classical trajectories. To validate the SC approach, we carried out the WD calculations and showed that transmission probabilities and classical trajectories are matching the SC ballistic predictions. The SC model takes advantage of representing the refraction index in terms of the wave vectors on both sides of the domain wall. This results in identical trajectories for the two considered domain walls whose transmission probabilities are indeed different. Within specific energy range, electrons can be highly collimated through the  considered system and steered by a magnetic field regardless the   types edges and domain walls. Most importantly, the considered system here is free of sharp electrostatic potential steps necessary for Klein tunneling and thus electron collimation. However, the major challenge in the experimental realization remains achieving SLG-AA domain walls which can be feasible in the near future as a result of the continued and decent development of graphene samples quality. Finally, we hope that our results will prove useful for designing graphene-based collimation optical devices that enable a new class of transport measurements. 
\begin{figure}[t!]
\centerline\centering\graphicspath{{./Figures/}}
\includegraphics[width=3 in]{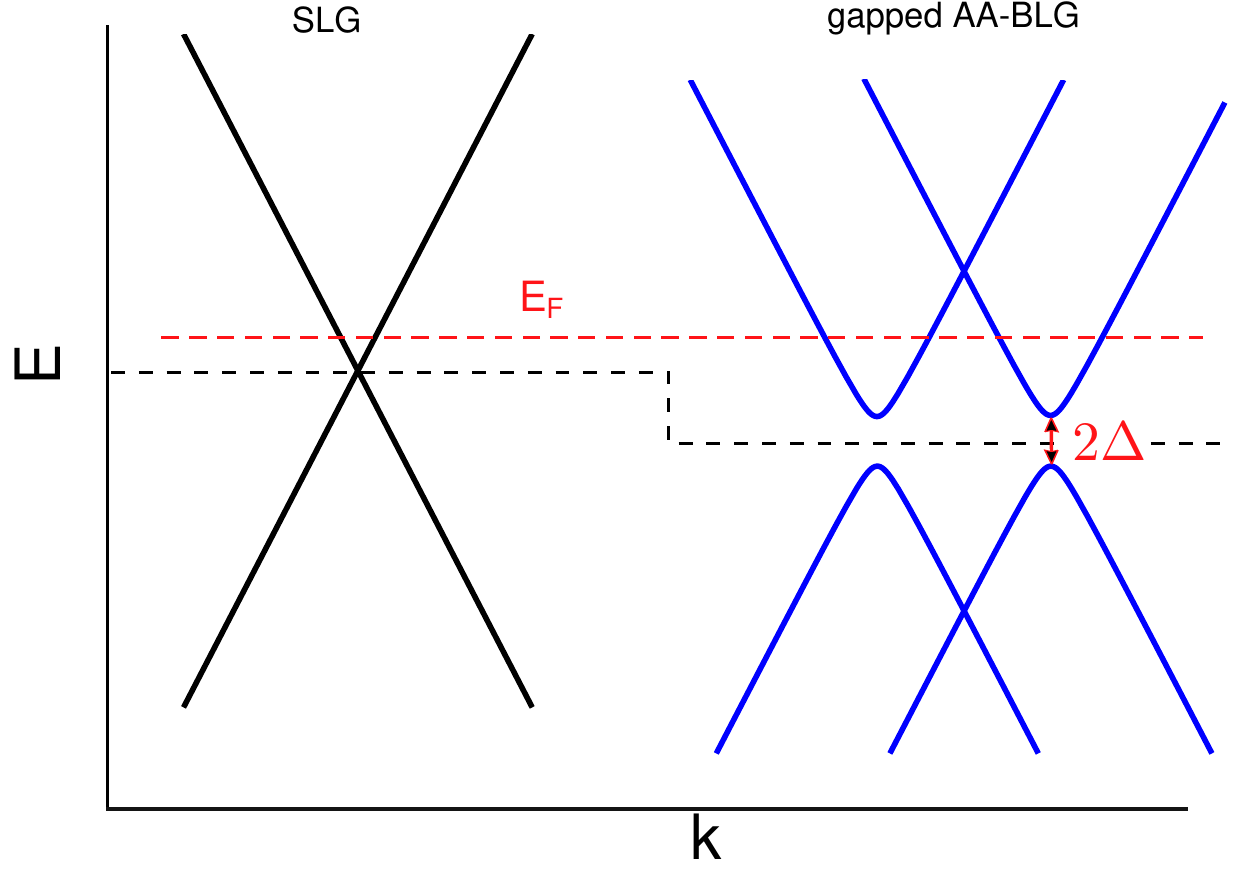}\\
\includegraphics[width=\linewidth]{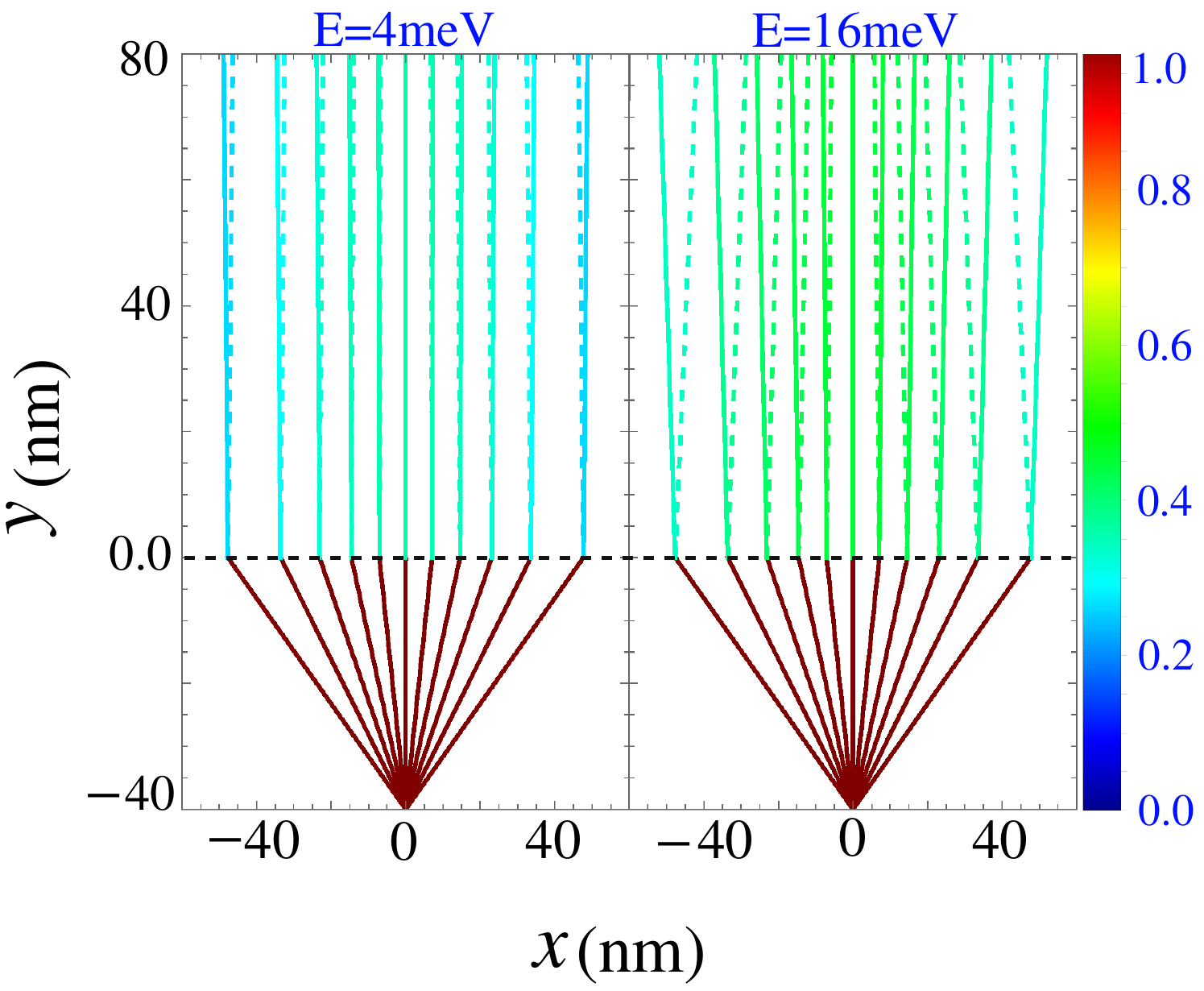}
\caption{(Color online) Top: energy bands of SLG and gapped AA\_BLG, dashed-black line represents the electrostatic potential $v_0$.  Bottom: SC electron collimation   as in Fig. \ref{fig:Traj_2SLG_AA} but for gAA with   $2\Delta=0.2\gamma_1$ and  $v_0=-2\Delta$. Note that for the other configurations the collimation is the same but the transmission probability will be slightly different.    } \label{gAA:transm}
\end{figure}
\begin{figure}[t!]
\centerline\centering\graphicspath{{./Figures/}}
\includegraphics[width=\linewidth]{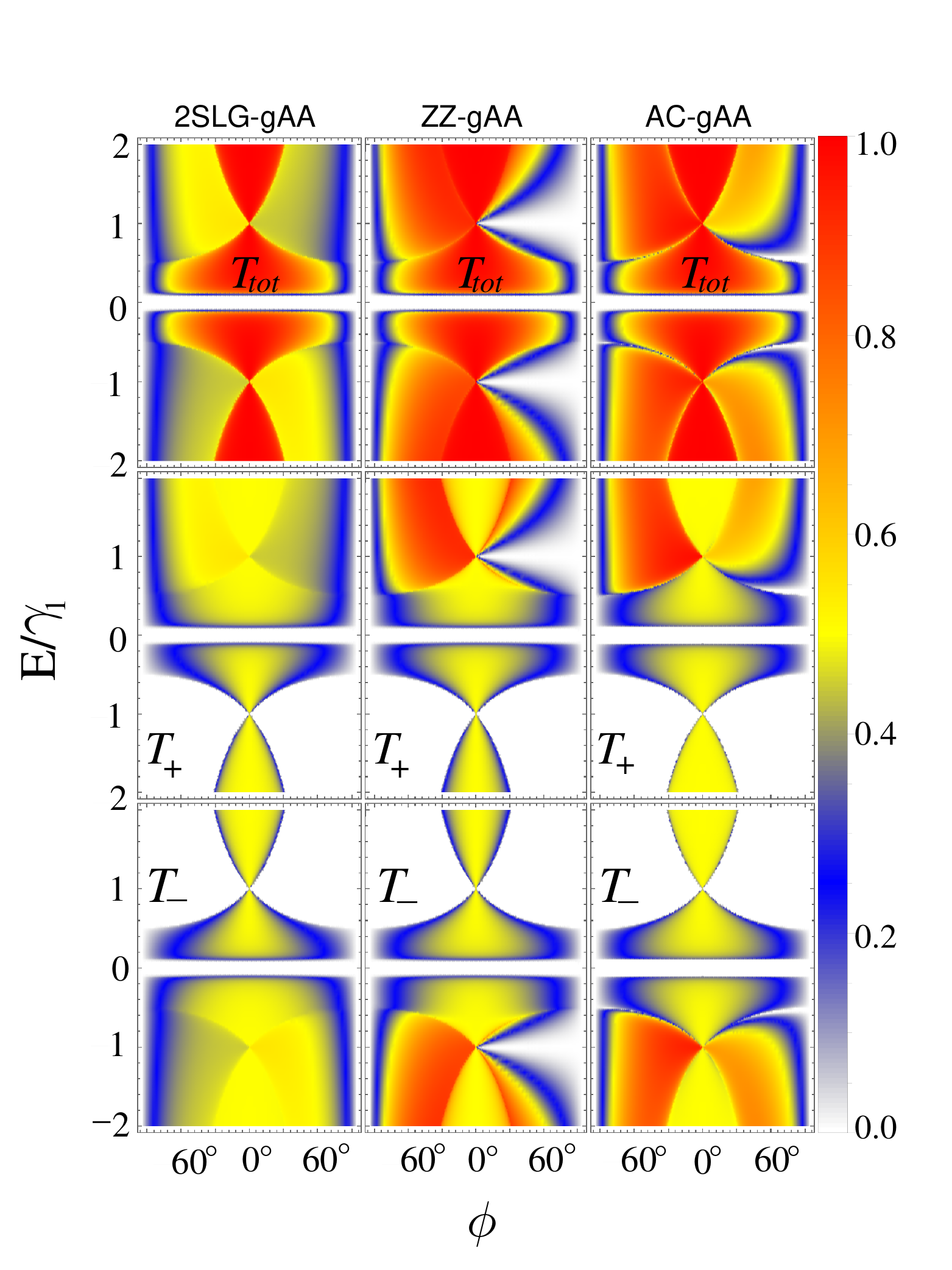}
\caption{(Color online) Cone and total  transmission  probabilities for gapped AA-BLG  for the three considered configurations  2SLG-gAA, ZZ-gAA, and AC-gAA as in Fig. \ref{SC_Transmission} with $v_0=0$ and a gap of magnitude $2\Delta=0.2\gamma_1$.  } \label{Transm_gAA}
\end{figure}
%
\section*{Acknowledgments}
H.M.A. and H.B.  acknowledge the support of King Fahd University
of Petroleum and Minerals under research group project No. RG181001. D.R.C and A.C were financially supported by the Brazilian Council for Research (CNPq) and CAPES foundation. BVD is supported by a postdoctoral fellowship by the Research Foundation Flanders (FWO-Vl).
\appendix 
\section{gapped AA-stacked bilayer graphene}\label{Append}
Through this paper we considered pristine AA-BLG whose spectrum is gapless and compose of two Dirac cones separated by $2\gamma_1$.  However, the more realistic spectrum is gapped due to the electron-electron interaction in graphene\cite{Rakhmanov_2012,Sboychakov_2013,Brey2013}. In this appendix we show that the electron collimation reported in this paper is maintained even in the presence of a finite gap in AA-BLG  energy spectrum.

In fact, the main effect of the gap coincides only with a slight change in the transmission probabilities. For the gapped AA-BLG (gAA), the continuum approximation for the Hamiltonian that
describes the electrons in the vicinity of the K-valley reads\cite{Abdullah2018c,Tabert2012}
\begin{equation}\label{gAA:H}
\mathcal{H}=\left(
\begin{array}{cccc}
  \Delta+v_0 & v_{F}\pi^{\dag} & \gamma_1 & 0 \\
  v_{F}\pi &  -\Delta+v_0&  0 & \gamma_1\\
  \gamma_1 &   0 &  -\Delta+v_0& v_{F}\pi^{\dag} \\
  0 & \gamma_1& v_{F}\pi & \Delta+v_0 \\
\end{array}%
\right).
\end{equation} 
The electron-electron interaction     breaks the layer and sublattice symmetries which results in the finite gap of magnitude  2$\Delta$ in the energy spectrum\cite{Rozhkov_2016}, see top panel of Fig. \ref{gAA:transm}. To investigate the collimation at the same Fermi energy considered in the case of pristine AA-BLG, we subject the gAA to an electrostatic gate  $v_0$ as presented by the dashed black line in the top panel of Fig. \ref{gAA:transm}. Then, we perform the same steps discussed in Sec. \ref{SCD} to calculate the electron collimation in gAA. In the bottom of Fig. \ref{gAA:transm} we show the electron collimation obtained using SC approach for  two different Fermi energies. We consider a gap of magnitude $2\Delta=0.2\gamma_1$ and an electrostatic gate of  strength $v_0=-2\Delta$. It appears that  the collimation is preserved even in  the presence of a finite energy gap, see Fig. \ref{fig:Traj_2SLG_AA} for comparison. This due to the fact that   the electron collimation is always preserved as long as the radius of the Fermi circle in AA-BLG region is much larger than its counterpart in SLG. Introducing the gap does not significantly alter the radius of the Fermi circle; however,  the  transmission probabilities are slightly reduced.
In Fig. \ref{Transm_gAA} we show the cone and total transmission probabilities in gAA for three configurations 2SLG-gAA, ZZ-gAA, and AC-gAA. As a comparison with  the results for pristine AA-BLG., we see that the transmission probability is drastically altered for energies around  the induced gap. It is completely suppressed  within the energy gap but apart from the gap and within the symmetric zones  the transmission probabilities are comparable for  both systems. Another difference is that as a results of braking the inversion symmetry the transmission probability for 2SLG-gAA is asymmetric with respect to normal incidence.  In conclusion, electron collimation can be preserved in both pristine and gapped AA-BLG, the only difference is that the latter one is not free  of electrostatic gate.

\begin{thebibliography}{10}

\bibitem{Cheianov2007}
V.~V. Cheianov, V.~Fal{\textquotesingle}ko, and B.~L. Altshuler,
\newblock Science {\bf 315}, 1252 (2007).

\bibitem{Banszerus2016}
L.~Banszerus, M.~Schmitz, S.~Engels, M.~Goldsche, K.~Watanabe, T.~Taniguchi,
  B.~Beschoten, and C.~Stampfer,
\newblock Nano Lett. {\bf 16}, 1387 (2016).

\bibitem{Wang2018a}
K.~Wang, M.~M. Elahi, K.~M.~M. Habib, T.~Taniguchi, K.~Watanabe, A.~W. Ghosh,
  G.-H. Lee, and P.~Kim,
\newblock http://arxiv.org/abs/1809.06757v2.

\bibitem{Sivan1990}
U.~Sivan, M.~Heiblum, C.~P. Umbach, and H.~Shtrikman,
\newblock Phys. Rev. B {\bf 41}, 7937 (1990).

\bibitem{Oliver1999}
W.~D. Oliver,
\newblock Science {\bf 284}, 299 (1999).

\bibitem{Hartmann2010}
R.~R. Hartmann, N.~J. Robinson, and M.~E. Portnoi,
\newblock Phys. Rev. B {\bf 81}, 245431 (2010).

\bibitem{Williams2011}
J.~R. Williams, T.~Low, M.~S. Lundstrom, and C.~M. Marcus,
\newblock Nat. Nanotechnol. {\bf 6}, 222 (2011).

\bibitem{Novoselov_2004}
K.~S. Novoselov, A.~K. Geim, S.~V. Morozov, D.~Jiang, Y.~Zhang, S.~V. Dubonos,
  I.~V. Grigorieva, and A.~A. Firsov,
\newblock Science {\bf 306}, 666 (2004).

\bibitem{Geim_2007}
A.~K. Geim, and K.~S. Novoselov,
\newblock Nat. Mater. {\bf 6}, 183 (2007).

\bibitem{Beenakker2008}
C.~W.~J. Beenakker,
\newblock Rev. Mod. Phys. {\bf 80}, 1337 (2008).

\bibitem{Klein_1929}
O.~Klein,
\newblock Zeitschrift f\"ur Physik {\bf 53}, 157 (1929).

\bibitem{Stander2009}
N.~Stander, B.~Huard, and D.~Goldhaber-Gordon,
\newblock Phys. Rev. Lett. {\bf 102}, 026807 (2009).

\bibitem{Katsnelson2006}
M.~I. Katsnelson, K.~S. Novoselov, and A.~K. Geim,
\newblock Nat. Phys. {\bf 2}, 620 (2006).

\bibitem{Gutierrez2016}
C.~Guti{\'{e}}rrez, L.~Brown, C.-J. Kim, J.~Park, and A.~N. Pasupathy,
\newblock Nat. Phys. {\bf 12}, 1069 (2016).

\bibitem{Abdullah2018a}
H.~M. Abdullah, and H.~Bahlouli,
\newblock J. Comput. Sci. {\bf 26}, 135 (2018).

\bibitem{Lee2015}
G.-H. Lee, G.-H. Park, and H.-J. Lee,
\newblock Nat. Phys. {\bf 11}, 925 (2015).

\bibitem{Chen2016}
S.~Chen, Z.~Han, M.~M. Elahi, K.~M.~M. Habib, L.~Wang, B.~Wen, Y.~Gao,
  T.~Taniguchi, K.~Watanabe, J.~Hone, A.~W. Ghosh, and C.~R. Dean,
\newblock Science {\bf 353}, 1522 (2016).

\bibitem{Veselago1968}
V.~G. Veselago,
\newblock Sov. Phys. Usp. {\bf 10}, 509 (1968).

\bibitem{Parimi2004}
P.~V. Parimi, W.~T. Lu, P.~Vodo, J.~Sokoloff, J.~S. Derov, and S.~Sridhar,
\newblock Phys. Rev. Lett. {\bf 92}, 127401 (2004).

\bibitem{Cubukcu2003}
E.~Cubukcu, K.~Aydin, E.~Ozbay, S.~Foteinopoulou, and C.~M. Soukoulis,
\newblock Phys. Rev. Lett. {\bf 91}, 207401 (2003).

\bibitem{Song2018}
J.~C.~W. Song, and N.~M. Gabor,
\newblock Nature Nanotechnology {\bf 13}, 986 (2018).

\bibitem{Houck2003}
A.~A. Houck, J.~B. Brock, and I.~L. Chuang,
\newblock Phys. Rev. Lett. {\bf 90}, 137401 (2003).

\bibitem{Grbic2004}
A.~Grbic, and G.~V. Eleftheriades,
\newblock Phys. Rev. Lett. {\bf 92}, 117403 (2004).

\bibitem{Aidala2007}
K.~E. Aidala, R.~E. Parrott, T.~Kramer, E.~J. Heller, R.~M. Westervelt, M.~P.
  Hanson, and A.~C. Gossard,
\newblock Nat. Phys. {\bf 3}, 464 (2007).

\bibitem{LaGasse2017}
S.~W. LaGasse, and J.~U. Lee,
\newblock Phys. Rev. B {\bf 95}, 155433 (2017).

\bibitem{Zhang2018}
S.-H. Zhang, W.~Yang, and F.~M. Peeters,
\newblock Phys. Rev. B {\bf 97}, 205437 (2018).

\bibitem{Sanderson_2013}
M.~Sanderson, Y.~S. Ang, and C.~Zhang,
\newblock Phys. Rev. B {\bf 88}, 245404 (2013).

\bibitem{Peterfalvi2012}
C.~G. P{\'{e}}terfalvi, L.~Oroszl{\'{a}}ny, C.~J. Lambert, and J.~Cserti,
\newblock New J. Phys. {\bf 14}, 063028 (2012).

\bibitem{Park2008}
C.-H. Park, Y.-W. Son, L.~Yang, M.~L. Cohen, and S.~G. Louie,
\newblock Nano Lett. {\bf 8}, 2920 (2008).

\bibitem{Choi2014}
S.~K. Choi, C.-H. Park, and S.~G. Louie,
\newblock Phys. Rev. Lett. {\bf 113}, 026802 (2014).

\bibitem{Liu2017a}
M.-H. Liu, C.~Gorini, and K.~Richter,
\newblock Phys. Rev. Lett. {\bf 118}, 066801 (2017).

\bibitem{Barnard2017}
A.~W. Barnard, A.~Hughes, A.~L. Sharpe, K.~Watanabe, T.~Taniguchi, and
  D.~Goldhaber-Gordon,
\newblock Nat. Commun. {\bf 8}, 15418 (2017).

\bibitem{Bhandari2018}
S.~Bhandari, G.~H. Lee, K.~Watanabe, T.~Taniguchi, P.~Kim, and R.~M.
  Westervelt,
\newblock 2D Mater. {\bf 5}, 021003 (2018).

\bibitem{Handschin2015}
C.~Handschin, B.~F\"ul\"op, P.~Makk, S.~Blanter, M.~Weiss, K.~Watanabe,
  T.~Taniguchi, S.~Csonka, and C.~Sch\"onenberger,
\newblock Appl. Phys. Lett. {\bf 107}, 183108 (2015).

\bibitem{Kinikar2017}
A.~Kinikar, T.~P. Sai, S.~Bhattacharyya, A.~Agarwala, T.~Biswas, S.~K. Sarker,
  H.~R. Krishnamurthy, M.~Jain, V.~B. Shenoy, and A.~Ghosh,
\newblock Nat. Nanotechnol. {\bf 12}, 564 (2017).

\bibitem{Overweg2017}
H.~Overweg, H.~Eggimann, X.~Chen, S.~Slizovskiy, M.~Eich, R.~Pisoni, Y.~Lee,
  P.~Rickhaus, K.~Watanabe, T.~Taniguchi, V.~Fal'ko, T.~Ihn, and K.~Ensslin,
\newblock Nano Lett. {\bf 18}, 553 (2017).

\bibitem{Boggild2017}
P.~B{\o}ggild, J.~M. Caridad, C.~Stampfer, G.~Calogero, N.~R. Papior, and
  M.~Brandbyge,
\newblock Nat. Commun. {\bf 8}, 15783 (2017).

\bibitem{Abdullah2017}
H.~M. Abdullah, B.~{\relax Van Duppen}, M.~Zarenia, H.~Bahlouli, and F.~M.
  Peeters,
\newblock J. Phys.: Condens. Matter {\bf 29}, 425303 (2017).

\bibitem{Abdullah2018}
H.~M. Abdullah, M.~{ Van der Donck}, H.~Bahlouli, F.~M. Peeters, and
  B.~{ Van Duppen},
\newblock Appl. Phys. Lett. {\bf 112}, 213101 (2018).

\bibitem{Abdullah2018b}
H.~M. Abdullah, H.~Bahlouli, F.~M. Peeters, and B.~{\relax Van Duppen},
\newblock J. Phys.: Condens. Matter {\bf 30}, 385301 (2018).

\bibitem{Abdullah_2016}
H.~M. Abdullah, M.~Zarenia, H.~Bahlouli, F.~M. Peeters, and B.~{ Van
  Duppen},
\newblock Europhys. Lett. {\bf 113}, 17006 (2016).

\bibitem{Lane2018}
T.~L.~M. Lane, M.~An{\dj}elkovi{\'{c}}, J.~R. Wallbank, L.~Covaci, F.~M.
  Peeters, and V.~I. Fal{\textquotesingle}ko,
\newblock Phys. Rev. B {\bf 97}, 045301 (2018).

\bibitem{Mirzakhani2016}
M.~Mirzakhani, M.~Zarenia, S.~A. Ketabi, D.~R. da~Costa, and F.~M. Peeters,
\newblock Phys. Rev. B {\bf 93}, 165410 (2016).

\bibitem{Reijnders2013}
K.~Reijnders, T.~Tudorovskiy, and M.~Katsnelson,
\newblock Ann. Phys. {\bf 333}, 155 (2013).

\bibitem{Milovanovic2015}
S.~P. Milovanovi{\'{c}}, D.~Moldovan, and F.~M. Peeters,
\newblock J. Appl. Phys. {\bf 118}, 154308 (2015).

\bibitem{Reijnders2017}
K.~J.~A. Reijnders, and M.~I. Katsnelson,
\newblock Phys. Rev. B {\bf 95}, 115310 (2017).

\bibitem{Milovanovic2014}
S.~P. Milovanovi{\'{c}}, M.~R. Masir, and F.~M. Peeters,
\newblock J. Appl. Phys. {\bf 115}, 043719 (2014).

\bibitem{Maksimova2008}
G.~M. Maksimova, V.~Y. Demikhovskii, and E.~V. Frolova,
\newblock Phys. Rev. B {\bf 78}, 235321 (2008).

\bibitem{Chaves2010}
A.~Chaves, L.~Covaci, K.~Y. Rakhimov, G.~A. Farias, and F.~M. Peeters,
\newblock Phys. Rev. B {\bf 82}, 205430 (2010).

\bibitem{Krueckl2009}
V.~Krueckl, and T.~Kramer,
\newblock New J. Phys. {\bf 11}, 093010 (2009).

\bibitem{Zhang_2011}
F.~Zhang, J.~Jung, G.~A. Fiete, Q.~Niu, and A.~H. MacDonald,
\newblock Phys. Rev. Lett. {\bf 106}, 156801 (2011).

\bibitem{Li2009}
Z.~Q. Li, E.~A. Henriksen, Z.~Jiang, Z.~Hao, M.~C. Martin, P.~Kim, H.~L.
  Stormer, and D.~N. Basov,
\newblock Phys. Rev. Lett. {\bf 102}, 037403 (2009).

\bibitem{Xu_2010}
Y.~Xu, X.~Li, and J.~Dong,
\newblock Nanotechnology {\bf 21}, 065711 (2010).

\bibitem{Lobato_2011}
I.~Lobato, and B.~Partoens,
\newblock Phys. Rev. B {\bf 83}, 165429 (2011).

\bibitem{Castro_Neto_2009}
A.~H. { Castro Neto}, F.~Guinea, N.~M.~R. Peres, K.~S. Novoselov, and
  A.~K. Geim,
\newblock Rev. Mod. Phys. {\bf 81}, 109 (2009).

\bibitem{Nakanishi_2010}
T.~Nakanishi, M.~Koshino, and T.~Ando,
\newblock Phys. Rev. B {\bf 82}, 125428 (2010).

\bibitem{Barbier01_2010}
M.~Barbier, P.~Vasilopoulos, and F.~M. Peeters,
\newblock Phys. Rev. B {\bf 82}, 235408 (2010).

\bibitem{Van_Duppen01_2013}
B.~{ Van Duppen}, and F.~M. Peeters,
\newblock Phys. Rev. B {\bf 87}, 205427 (2013).

\bibitem{Abdullah_2017}
H.~M. Abdullah, A.~E. Mouhafid, H.~Bahlouli, and A.~Jellal,
\newblock Mater. Res. Express {\bf 4}, 025009 (2017).

\bibitem{Pereira2010}
J.~M. Pereira, F.~M. Peeters, A.~Chaves, and G.~A. Farias,
\newblock Semicond. Sci. Technol. {\bf 25}, 033002 (2010).

\bibitem{Masirz2010}
M.~R. Masir, P.~Vasilopoulos, and F.~M. Peeters,
\newblock Phys. Rev. B {\bf 82}, 115417 (2010).

\bibitem{Barbier2010}
M.~Barbier, P.~Vasilopoulos, and F.~M. Peeters,
\newblock Phil. Trans. R. Soc. A {\bf 368}, 5499 (2010).

\bibitem{Park2011}
S.~Park, and H.-S. Sim,
\newblock Phys. Rev. B {\bf 84}, 235432 (2011).

\bibitem{Phong2016}
V.~T. Phong, and J.~F. Kong,
\newblock arXiv: 1610.00201v1  (2016).

\bibitem{Ariel2013}
V.~Ariel, and A.~Natan,
\newblock Electron effective mass in graphene,
\newblock in {\em 2013 International Conference on Electromagnetics in Advanced
  Applications ({ICEAA})}, {IEEE}, 2013.

\bibitem{Ashcroft1976}
N.~W. Ashcroft, and N.~D. Mermin,
\newblock {\em Solid State Physics} (Cengage Learning, Boston, 1976), pp.
  231--233.

\bibitem{Zou2011}
K.~Zou, X.~Hong, and J.~Zhu,
\newblock Phys. Rev. B {\bf 84}, 085408 (2011).

\bibitem{Batista2018}
F.~Batista, A.~Chaves, D.~R. da~Costa, and G.~Farias,
\newblock Physica E {\bf 99}, 304 (2018).

\bibitem{Costa2017}
D.~R. da~Costa, A.~Chaves, G.~A. Farias, and F.~M. Peeters,
\newblock J. Phys.: Condens. Matter {\bf 29}, 215502 (2017).

\bibitem{Chaves2015a}
A.~Chaves, D.~R. da~Costa, G.~O. de~Sousa, J.~M. Pereira, and G.~A. Farias,
\newblock Phys. Rev. B {\bf 92}, 125441 (2015).

\bibitem{Cavalcante2016}
L.~S. Cavalcante, A.~Chaves, D.~R. da~Costa, G.~A. Farias, and F.~M. Peeters,
\newblock Phys. Rev. B {\bf 94}, 075432 (2016).

\bibitem{Costa2012}
D.~R. da~Costa, A.~Chaves, G.~A. Farias, L.~Covaci, and F.~M. Peeters,
\newblock Phys. Rev. B {\bf 86}, 115434 (2012).

\bibitem{da_Costa_2015}
D.~R. da~Costa, A.~Chaves, S.~H.~R. Sena, G.~A. Farias, and F.~M. Peeters,
\newblock Phys. Rev. B {\bf 92}, 045417 (2015).

\bibitem{Chaves2009}
A.~Chaves, G.~A. Farias, F.~M. Peeters, and B.~Szafran,
\newblock Phys. Rev. B {\bf 80}, 125331 (2009).

\bibitem{Degani2010}
M.~H. Degani, and M.~Z. Maialle,
\newblock J. Comput. Theor. Nanosci. {\bf 7}, 454 (2010).

\bibitem{Chaves2015}
A.~Chaves, G.~A. Farias, F.~M. Peeters, and R.~Ferreira,
\newblock Comm. Comput. Phys. {\bf 17}, 850 (2015).

\bibitem{Rakhimov2011}
K.~Y. Rakhimov, A.~Chaves, G.~A. Farias, and F.~M. Peeters,
\newblock J. Phys.: Condens. Matter {\bf 23}, 275801 (2011).

\bibitem{Kramer2011}
T.~Kramer,
\newblock AIP Conf. Proc. {\bf 1334}, 142 (2011).

\bibitem{Yan2016}
W.~Yan, S.-Y. Li, L.-J. Yin, J.-B. Qiao, J.-C. Nie, and L.~He,
\newblock Phys. Rev. B {\bf 93}, 195408 (2016).

\bibitem{Clark_2014}
K.~W. Clark, X.-G. Zhang, G.~Gu, J.~Park, G.~He, R. M.~Feenstra, and A.-P. Li,
\newblock Phys. Rev. X {\bf 4}, 011021 (2014).

\bibitem{Rakhmanov_2012}
A.~L. Rakhmanov, A.~V. Rozhkov, A.~O. Sboychakov, and F.~Nori,
\newblock Phys. Rev. Lett. {\bf 109}, 206801 (2012).

\bibitem{Sboychakov_2013}
A.~O. Sboychakov, A.~L. Rakhmanov, A.~V. Rozhkov, and F.~Nori,
\newblock Phys. Rev. B {\bf 87}, 121401(R) (2013).

\bibitem{Brey2013}
L.~Brey, and H.~A. Fertig,
\newblock Phys. Rev. B {\bf 87}, 115411 (2013).

\bibitem{Abdullah2018c}
H.~M. Abdullah, M.~A. Ezzi, and H.~Bahlouli,
\newblock J. Appl. Phys. {\bf 124}, 204303 (2018).

\bibitem{Tabert2012}
C.~J. Tabert, and E.~J. Nicol,
\newblock Phys. Rev. B {\bf 86}, 075439 (2012).

\bibitem{Rozhkov_2016}
A.~Rozhkov, A.~Sboychakov, A.~Rakhmanov, and F.~Nori,
\newblock Phys. Rep. {\bf 648}, 1 (2016).

\end{thebibliography}


\end{document}